\documentclass[showkeys,preprint,twocolumn,11pt,superscriptaddress,floatfix]{revtex4-2}
\usepackage[utf8]{inputenc}
\usepackage[english]{babel}
\usepackage{graphicx}
\usepackage{dcolumn} % Align table columns on decimal point
\usepackage{bm} % bold math
\usepackage[colorlinks=true,
            linkcolor=blue,
            urlcolor=blue,
            citecolor=blue]{hyperref}
\usepackage[mathlines]{lineno} % Enable numbering of text and display math
\usepackage{amsmath, amsfonts}
\usepackage{physics}
\usepackage{booktabs}
\usepackage{tabularx}
\usepackage{geometry}
\begin{document}

\title{Route to chaos and chimera states in a network of memristive Hindmarsh-Rose neuron model with external excitation}
\author{Sishu Shankar Muni}
%\ead{sishushankarmuni@gmail.com}
\affiliation{Department of Physical Sciences, Indian Institute of Science Education and Research Kolkata, Mohanpur, West Bengal, 741246, India}

\author{Zeric Tabekoueng Njitacke}
%\email{vagdsantos@gmail.com}
\affiliation{Department of Automation, Biomechanics and Mechatronics\\ Lodz University of Technology, Lodz, Poland.}
\affiliation{Research Unit of Automation and Applied Computer (URAIA), Electrical Engineering Department of IUT-FV, University of Dschang, P.O. Box 134, Bandjoun, Cameroon.}
\affiliation{Department of Electrical and Electronic Engineering, College of Technology (COT), University of Buea, P.O.Box 63, Buea, Cameroon}
\author{Cyrille Feudjio}
%\email{matheusrolim95@gmail.com}
\affiliation{Department of Electrical and Electronic Engineering, College of Technology (COT), University of Buea, P.O.Box 63, Buea, Cameroon.}
\author{Theophile Fozin}
\affiliation{Department of Electrical and Electronic Engineering, College of Technology (COT), University of Buea, P.O.Box 63, Buea, Cameroon}
\author{Jan Awrejcewicz}
%\ead{jdsjunior@uepg.br}    
\affiliation{Department of Automation, Biomechanics and Mechatronics\\ Lodz University of Technology, Lodz, Poland.}

\begin{abstract}
In this paper we have introduced and investigated the collective behavior of a network of memristive Hindmarsh-Rose (HR) neurons. The proposed model was built considering the memristive autapse of the traditional 2D HR neuron. Using the one-parameter bifurcation diagram and its corresponding maximal Lyapunov exponent graph, we showed that the proposed model was able to exhibit a reverse period doubling route to chaos, phenomenon of interior and exterior crises. Three different configurations of the ring-star network of the memristive HR neuron model, including ring-star, ring, and star, have been considered. The study of those network configurations revealed incoherent, coherent , chimera and cluster state behaviors. Coherent behavior is characterized by synchronization of the neurons of the network, while incoherent behaviors are characterized by the absence of synchronization. Chimera states refer to a differet state where
there is a coexistence of synchronized and asynchronized nodes of the network. One of the interesting result of the paper is the prevalence of double-well chimera states in both ring and ring-star network and has been first mentioned in the case of memrisitve HR neuron model.
\end{abstract}

\maketitle
%==========================================================    
%====================== INTRODUCTION ======================    
%==========================================================    
\section{\label{sec:level1}Introduction}
A brain, like a complex organ, is built from the interconnection of a very large number of neurons. These interconnected neurons are very important because they are the seat of the processing, calculation, storage, and transfer of information \citep{ref1}. These neurons are connected to each other using a synapse. As a result, a synapse is the part of the nervous system that allows a presynaptic neuron the transmission of electrical or chemical signals to the postsynaptic neuron \citep{ref2}.  As a result, several mathematical models have been developed and studied in the literature to study some of the dynamical mechanisms of the brain. 
The Hopfield neural network model \citep{ref3,ref4,ref5,ref6}, the Hodgkin-Huxley neuron  \citep{ref7}, the 2-D Hindmarsh-Rose (HR), the 3D-HR neuron models \citep{ref12,ref13}, the FitzHugh-Nagumo (FHN) neuron model \citep{ref10, korkmaz4fpga}, the MorrisLecar neuron model \citep{ref11}, the Chay neuron model \citep{ref8}, the Izhikevich neuron model \citep{ref9}, and the Rulkov neuron model \citep{ref14} are some examples.
 In the same vein, several artificial synapse models for presynaptic and postsynaptic neuron coupling have been developed in the literature. Some of them are   hybrid synapse \citep{ref18}, Josephson junction synapse \citep{ref19}, memristive synapse \citep{ref20}, electrical synapse \citep{ref16,ref17}, and  chemical synapse \citep{ref15}.  Following that, a large number of single neurons \citep{ref21,ref22,ref23,ref24,ref25,ref26,ref27,ref28} and coupled neurons \citep{ref26,ref29,ref30,ref31,ref32,ref33,ref34,ref35,ref36,ref37,ref38,ref39,ref40,ref41}  models have been introduced and addressed in the literature using the quoted artificial synapses. 

The authors of \citep{qin2021phase} investigated the phenomenon of phase-amplitude coupling in nonlinearly coupled Stuart-Landau oscillators. Among the architectures used by the authors, it can be found the high-frequency neural oscillation driven by an external low-frequency input and two interacting local oscillations with distinct, locally generated frequencies.  The problem of reconstructing the model equations for the network of 3rd order neuron-like oscillators from time series has been addressed in ref. \citep{sysoeva2021reconstruction}. The authors showed that by using phase-locked loop systems as nodes of the networks, dynamical regimes such as quasiharmonic oscillations, spiking, bursting, and chaotic behavior are based on different network typologies such as star, ring, chain, and random architectures. The dynamical and physiological effects of the presence of electric field on an improved version of FitzHugh-Nagumo model was investigated in  \citep{takembo2022modulational}. Using the multiple scale expansion method on the system of N-differential equations, the authors obtained the angular frequency of the modulated impulse wave along the network. Finally, the formation of localized nonlinear wave patterns was confirmed in the proposed network.
The behavior of both single and a network of FHN neuron with memristors were investigated in  ref. \citep{njitacke2022hamilton}. The investigation of the single neuron revealed the presence of hidden dynamics, which is an interesting feature in the qualitative theory of dynamical systems. The biophysical energy of that model was established using the famous the Helmholtz theorem. The authors found that variation of external current on the model had no effect on the energy. Interestingly, the autapse coupling strength affects the energy released by the neuron. A plethora of spiking and bursting patterns is observed in the model. Hysteretic dynamics due to the coexistence of different firing patterns was confirmed.  To verify both the analytical, numerical results, an equivalent electronic circuit was constructed. It was found that the results obtained from the circuit are in good agreement with the numerical simulations.  In the end, information pattern stability was explored statistically via modulational instability under memristive autapse strength using a chain network of 500 identical neurons. It was discovered that the new network enables localized information patterns with attributes of synchronization as a means of information coding when initial conditions are considered as slightly modulated plane waves. The improved information coding pattern and potential mode transition were also confirmed by stronger autaptic couplings caused by fixing the stimulation current. 

\textcolor{black}{After researchers have studied coupled pendulums and their dynamical behavior \cite{Huy17}, there has been a plethora of studies on network of oscillators. When the  network elements have similar phases and frequencies the oscillators get synchronized. If the phases and frequencies are different they get desynchronized. Kuramoto found a new type of network state in which oscillators synchronize and desynchronize in a network of oscillators and these were termed as chimera states \cite{kuramoto2002coexistence}. There has been a lot of works on chimeras thereafter \cite{scholl2016synchronization, majhi2019chimera, omel2018mathematics, panaggio2015chimera}, just to name a few.   Scientists have even uncovered epilepsy and schizophernia as topological diseases that depend on the topology of the neurons interconnected in the brain \cite{uhlhaas2006neural}. Neurons can also be considered as dynamical oscillators and in brain millions of neurons are interconnected in a complex fashion and neurons transmit nerve signals and sensory informations.  This motivates to study the behavior of networks in neuron oscillators. }

  An autapse is a specific synapse developed from an auxiliary loop that enables it to connect the axon and the dendrite of the same neuron together. In this contribution, a memristor is introduced in a 2D Hindmarsh-Rose neuron model. Therefore, the memristive Hindmarsh-Rose neuron thus obtained is also called the 2D Hindmarsh-Rose neuron with a memristive autapse. The study of the network is based on the ring-star, ring, and star connection from the introduced model. So the  outline of the paper is as follows: In Section 2, the mathematical model of the memristive Hindmarsh-Rose model is discussed. Its complex dynamical behavior is revealed through some numerical simulations. Its network topology is also presented. In Section 3, numerical simulations are used to explore the collective behavior of the various network topologies considered. Lastly, in Section 4, we conclude and present scope for further research work. All the simulations in the paper is carried out using \textsc{Matlab}.

		\section{Presentation of the neuron model }
	
\subsection{Framework of memristive autapse }
When an axon is injured, such as by poisoning in ion channels or heterogeneity in a local area of the axon, signal transmission can be terminated or blocked during neuronal communication. As a result, neurons can develop new loops or secondary loops to help with signal transmission. This auxiliary loop is known as an autapse, which can be electrical autapse current, chemical autapse current, or memristive autapse current. 
 Using memristor definition \citep{ref48} and applying Ohm's law, we get Eq.(\ref{eq1}).The term $G(u)$ represents the memductance and $ i, u, v $ are state variables.
   \begin{equation}\label{eq1}
 	\left\{ \begin{array}{l}
 	 i_m  = G\left( u \right)v = \alpha \cos \left( u \right)v, \\ 
 	 \frac{{du}}{{dt}} = g\left( {u,v} \right) = \sin \left( u \right) + ev. \\ 
 	 \end{array} \right.
 	 \end{equation}
 	 
 	 \begin{figure}[htbp]
 	 	\begin{center}
 	 		\includegraphics[width=0.23\textwidth]{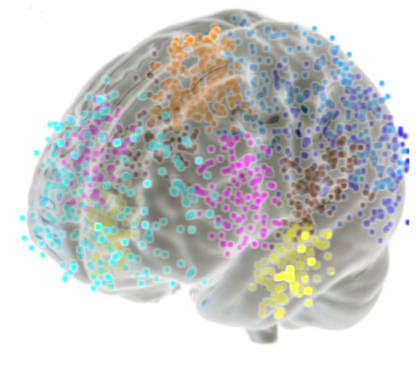} 
 	 	\end{center}
 	 	\caption{\textcolor{black}{Complex interconnections of millions of neurons in the brain\citep{ref48}}}\label{fig1} 
 	 \end{figure}
	The memristive nature of the autapse proposed in (\ref{eq1}) is supported by the well-known fingerprint of the memristor, characterized by a pinched hysteresis loop at the origin of the current-voltage characteristic when applying an external stimulus in the form $v = A\sin \left( {Ft} \right)$. For the sake of brevity, that result is not provided.  \textcolor{black}{Recall that the memristive neuron model used for this investigation was previously introduced in \citep{njitacke2022energy}. In that work, the global dynamical behavior as well as the effect of the initial condition on the behavior of the neuronal model have been investigated. The authors discovered the considered neuron model with memristive autapse was able to exhibit a homogenous extreme multistability characterized by the coexistence of an infinite number of patterns of the same shape. Since the work was focused only on the dynamics of a single neuron, the investigation of the collective behavior of such a model with homogeneous extreme multistability has further supported the aim of this study.}
\subsection{Design of the coupled neurons}
Neurons are the central organs of the brain since they enable computation, processing, and storage of information, just to name a few. As it can be seen in Fig.(\ref{fig1}), the brain is made up of interconnections of a very large number of neurons. As a result, the investigation of a ring-star network of  neurons composed of Hindmarsh-Rose neurons with memristive autapse will be addressed in this contribution. The mathematical model of the memristive HR neuron   is given in (\ref{eq2}).

 \begin{equation}\label{eq2}
 	\left\{ \begin{array}{l}
 	 \dot x = y - ax^3  + bx^2  + \alpha \cos \left( u \right)x + i_s  \\ 
 	 \dot y = c - dx^2  - y \\ 
 	 \dot u = \sin \left( u \right) + ex \\ 
 	 \end{array} \right.
  \end{equation}

\begin{figure}[!t]
	\centering
	\includegraphics[width=0.23\textwidth]{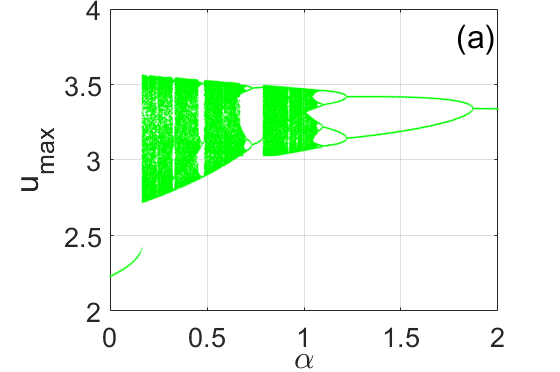} 
	\includegraphics[width=0.23\textwidth]{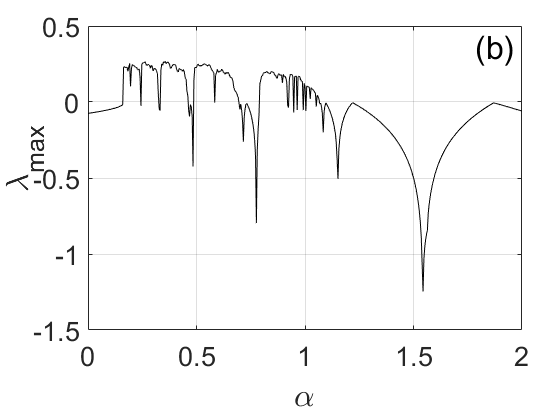}
	\caption{In (a), one-parameter bifurcation diagram of $u$ vs $\alpha$ showing reverse period-doubling route to \textcolor{black}{chaos  as parameter $\alpha$ is increased.} In (b), the corresponding Lyapunov exponent is estimated numerically. The parameters are set as $f=0.5$, $m=2$ with the initial condition $(0,0,1)$. } 
	\label{fig2}
\end{figure}

\begin{figure}[!t]
	\centering
	\includegraphics[width=0.23\textwidth]{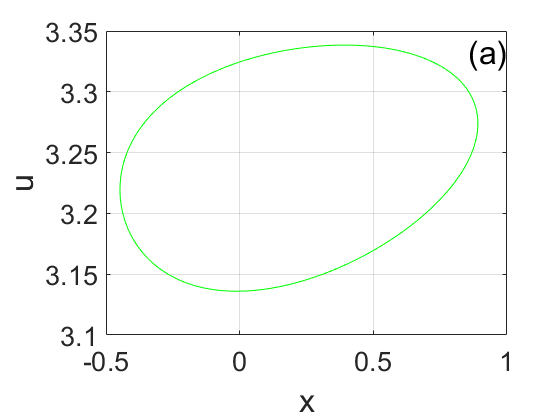} 
	\includegraphics[width=0.23\textwidth]{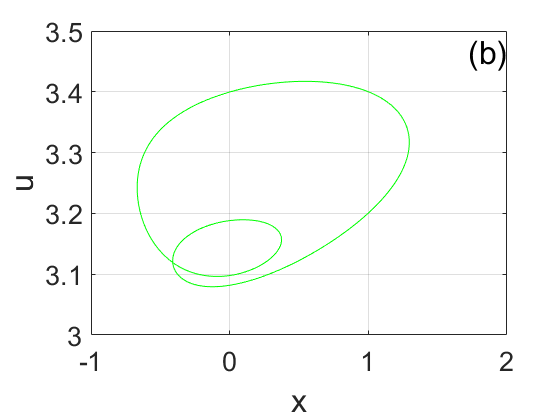}
		\includegraphics[width=0.23\textwidth]{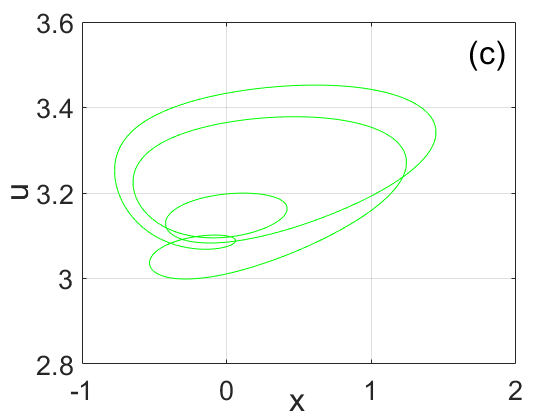} 
	\includegraphics[width=0.23\textwidth]{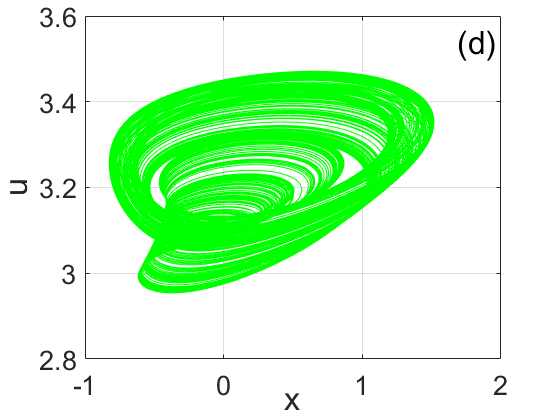}
	\caption{\textcolor{black}{Phase space trajectories showing the phenomenon of the reverse period doubling bifurcation for some discrete values of the control parameter $ \alpha $ } }
	\label{fig3a}
\end{figure}

\begin{figure}[!t]
\centering
 \includegraphics[width=0.23\textwidth]{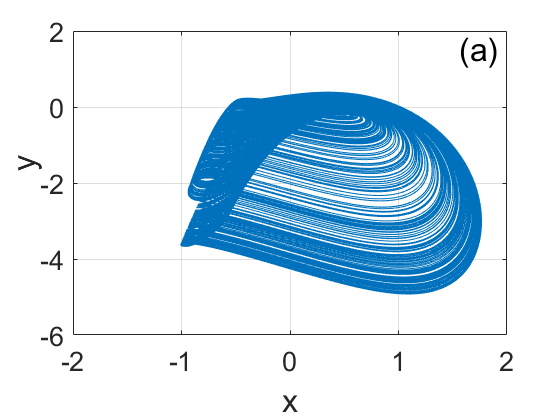} 
  \includegraphics[width=0.23\textwidth]{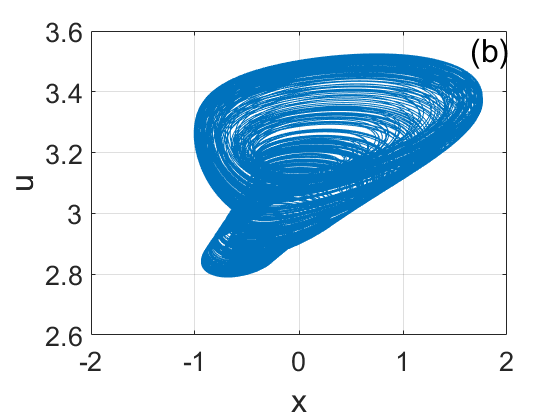}
  \includegraphics[width=0.23\textwidth]{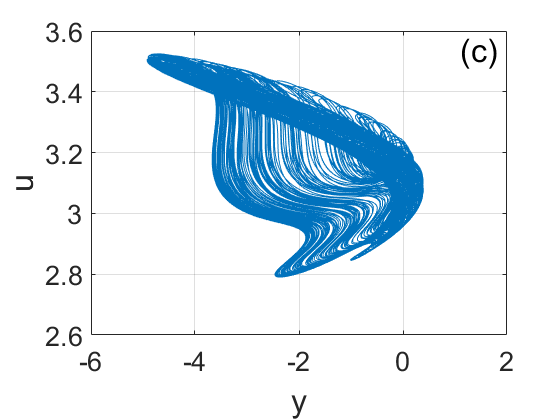}
   \includegraphics[width=0.23\textwidth]{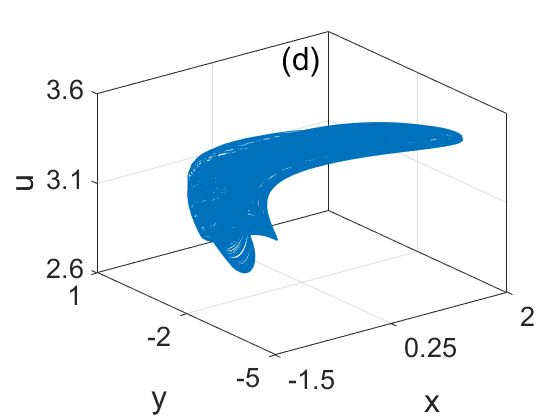}
    \caption{Phase space trajectories for a discrete value $\alpha=0.5$ of model of the neuron memristive  autapse displaying chaotic dynamics.} \label{fig3}
\end{figure}

In (\ref{eq2}), $x$   is the membrane potential of the HR neuron, $y$ represents the retrieval variable related to a fast current of either $Na^{+}$ or $K^{+}$. The state variable $u$ stands for the inner variable of the memristive autapse , variable $i_s  = m\sin \left( {2\pi ft} \right)$ represents outward input  current and $ \alpha $ indicates the connection strength of the memristive autapse. For parameters $a=1, b=3, c=1, d=5, e=0.5, m=2, f=0.5$ and $\alpha$ is tuneable.
As it can be seen in Fig.(\ref{fig2}).  The single HR neuron with a memristive autapse is able to exhibit very rich and striking bifurcations. When decreasing the control parameter, phenomena such as reverse period doubling bifurcation, interior and exterior crises are observed. These crises occur when a chaotic motion is suddenly destroyed and gives birth to periodic motion, or when a chaotic motion is suddenly created from a periodic one instead of being destroyed.  \textcolor{black}{ As it can be seen in Fig.(\ref{fig3a}), four phase space trajectories have been computed to further support the phenomenon of the reverse period doubling bifurcation when the memristive autapse strength $ \alpha  $ is decreased. When decreasing $ \alpha  $, it is observed that period-1 for $ \alpha=2 $, period-2 for  $ \alpha=1.5 $, period-4 for $ \alpha=1.15 $ and a chaotic attractor for $ \alpha=1 $.}
In addition, as the control parameter in the system is varied, alternating transitions of periodic and chaotic behavior is observed. When the control parameter $\alpha$ is tuned to $0.5$, some two-dimensional and three-dimensional projections of the chaotic activity, generated by the memristive neuron of the network considered in this work, are provided in Fig.(\ref{fig3}).

\section{Ring-star network of memristive Hindmarsh-Rose model}
\begin{figure}
	\centering
	\includegraphics{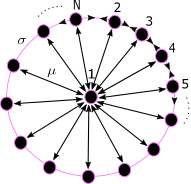}
	\caption{The memristive Hindmarsh-Rose neuron system connected in a ring-star network. Here we consider $N=100$ \textcolor{black}{memristive HR neurons where the central one is labeled $i=1$ and the end nodes are labeled from $i=2,\ldots,N$. The ring and star coupling strengths are denoted by $\sigma$ and $\mu$ respectively.}}
	\label{networkfig1}
\end{figure}

After exploring the dynamical analysis of the memristive Hindmarsh-Rose neuron model in brief, we explore the collective behavior in a ring-star network of memristive Hindmarsh-Rose neuron model. An advantage of this mixed topological network is we get three different networks for free (ring-star, ring, and star network). Ring and star networks find their applications in various real world systems \citep{ref49}, gene regulatory networks \citep{ref50}, just to name a few. It is an important fundamental network to study first for a dynamical oscillator.
A sketch of a ring-star network is illustrated in Fig. \ref{networkfig1}.  
\begin{figure*}[!htbp]
	\begin{center}
		\includegraphics[width=0.90\textwidth]{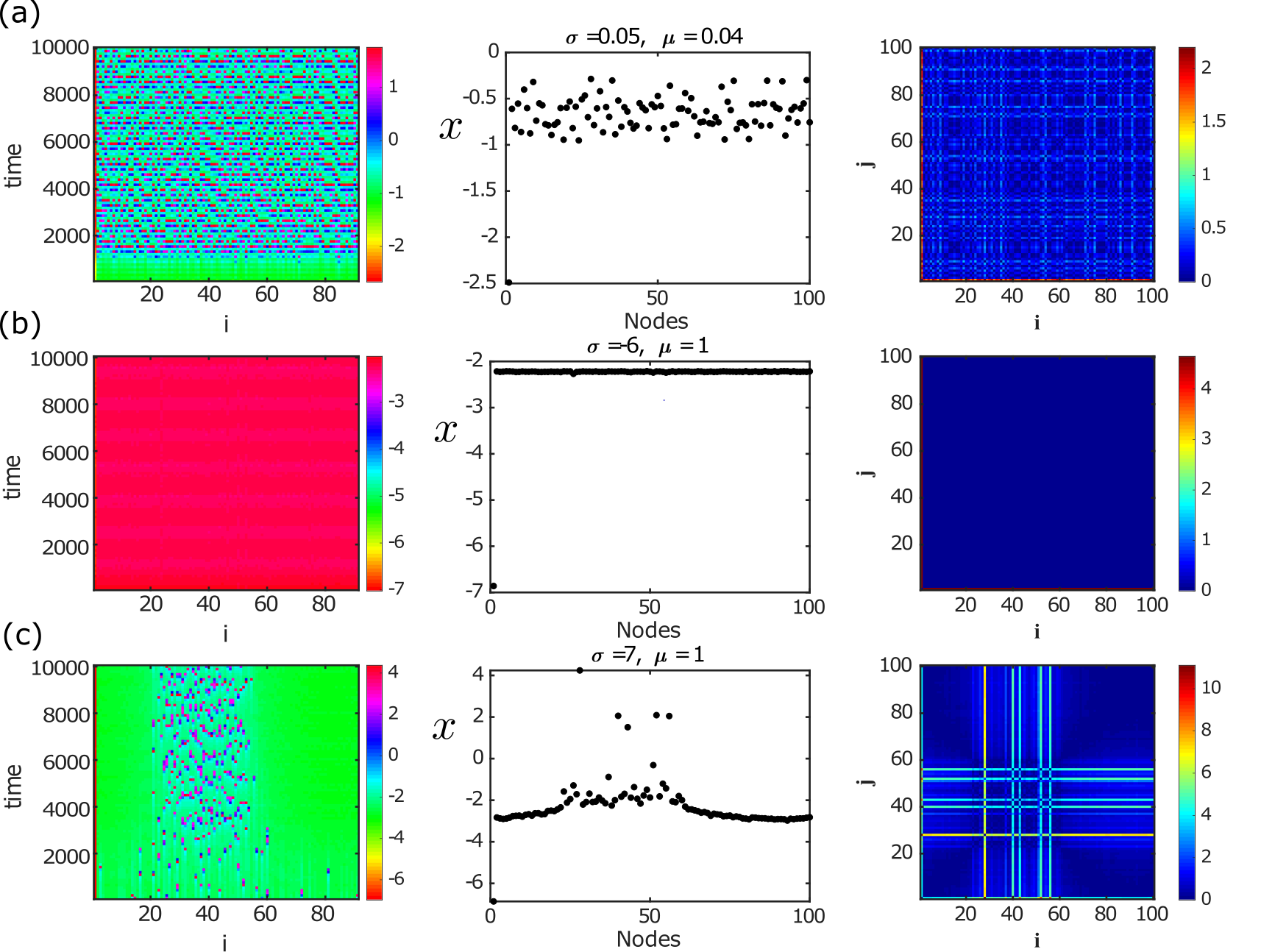}
	\end{center}
	\caption{Ring-Star network of memristive Hindmarsh Rose neuron model with $\sigma \neq 0, \mu \neq 0$. Random initial conditions are set with the coupling range of $P =70$. Number of nodes considered is $N =100$. Asynchronous behavior in (a), synchronous behavior in (b), chimera state in (c) is shown.}
	\label{fig:RingStarone}
\end{figure*}

The dynamical equations of the ring-star network are given by
\begin{align}
	\begin{split}
		\label{eq:diffusive}
		\dot{x_{i}} &= f_{x} + \mu(x_{i}-x_{1}) + \frac{\sigma}{2P} \sum_{n=i-P}^{n=i+P}(x_{i}-x_{n}),\\
		\dot{y_{i}} &= f_{y},\\
		\dot{u_{i}} &= f_{u},\\
		\dot{w_{i}} &= f_{w}.
	\end{split}
\end{align}
\noindent The central node ($i=1$) is governed by the following system of differential equations: 
\begin{align}
	\begin{split}
		\label{eq:central}
		\dot{x_{1}} &= f_{x} + \sum_{j=1}^{N} \mu(x_{j} - x_{1}),\\
		\dot{y_{1}} &= f_{y},\\
		\dot{u_{1}} &= f_{u},\\
		\dot{w_{1}} &= f_{w}.
	\end{split}
\end{align}
\noindent where
\begin{align*}
	f_{x} &= y_{i} - ax_{i}^3 + bx_{i}^2 - \alpha x_{i} \cos(u_{i}) + m \sin(w_{i}),\\
	f_{y} &= c - dx_{i}^2 - y_{i},\\
	f_{u} &= \sin(u_{i}) + ex_{i},\\
	f_{w} &= 2\pi f.
\end{align*}
\noindent with periodic boundary conditions:
\begin{align*} 
	x_{i+N}(t) &= x_{i}(t),\\
	y_{i+N}(t) &= y_{i}(t),\\
	u_{i+N}(t) &= u_{i}(t),\\
	w_{i+N}(t) &= w_{i}(t)
\end{align*}
\noindent for $i=2,3,\ldots,N$.
The parameters used throughout this study are: $a = 1, b = 3, c = 1, d = 5, e = 0.5, \alpha = 0.5, f = 0.5$. The size of the network is considered to be of $100$ nodes with $P$ nearest neighbors connected to each other. The network parameters such as the ring coupling strength $\sigma$, star coupling strength $\mu$, and the coupling range $P$ will be varied to explore different synchronization patterns arising in the ring-star network of memristive Hindmarsh-Rose neuron system.

Note that the ring-star network transforms to a ring network when $\mu=0$ and it transforms to a star network when $\sigma = 0$. The mixed topological ring-star network prevails when $\sigma \neq 0$ and $\mu \neq 0$. We have divided our whole network analysis into three categories: category A: ring-star network, category B: ring network, category C: star network.

\subsection{Characterization of chimera states}
In order to characterize the spatiotemporal patterns obtained in the study, we use the measure of strength of incoherence (SI). SI was developed as a measure to characterize different spatiotemporal patterns exhibited by the network of neurons. Many studies have shown that SI is able to characterize different spatiotemporal states in a network of neurons.

Here we give  a sketch of the method adapted by the Strength of Incoherence. The idea lies in transforming the original variables into new variables. Suppose $x_{i}, i = 1, \ldots , N$ represents the original set of variables of the network system. Next, define new set of variables as $z_{i} = x_{i} - x_{i-1}, i  =1, \ldots, N$. The average of $z_{i}$'s is denoted by \begin{equation*}
    \langle z \rangle = \frac{1}{N} \sum_{i=1}^{N} z_{i}.
\end{equation*} 
We then evaluate the quantity 
\begin{equation*}
\chi(m) = \biggl< \sqrt{\sum_{j=n(m-1)+1}^{nm} (z_{j} - \langle z\rangle)^2}\rangle_{t} \biggr>. 
\end{equation*} We calculate $s_{m} = \Theta(\delta - \chi(m))$, where $\delta$ is a predetermined threshold based on which different characterizations of the network is carried out, and $m$ denotes the number of bins the network is grouped, $m = N/n$. Strength of Incoherence (SI) is then defined as \begin{equation}\rm{SI} = 1 - \frac{\sum_{m=1}^{M} s_{m}}{M}\end{equation}
If $\rm{SI} \approx 1$, it denotes incoherent state, if $\rm{SI} = 0 $, it denotes a synchronized state, cluster state, and if $0<\rm{SI}<1$, it denotes chimera state.   

\subsection{Ring-star network}
\label{sec:ringstarnet}
Here we consider the effect of both ring and star coupling strengths $(\sigma \neq 0, \mu \neq 0)$ for our network and analyse the spatiotemporal patterns. 
When $\sigma = 0.05, \mu = 0.04$, the neuron nodes exhibit asynchronous patterns, see Fig. \ref{fig:RingStarone} (a) showing the nodes oscillating in an asynchronous fashion. The SI value is also $1$, signifying incoherence.  The leftmost plot shows the variation of the membrane potential ($x$) as time evolves.  The right most plot illustrates the recurrence plot of the nodes of the network under study by considering the Euclidean norm of the $x$ state values of different nodes. Each point $(i,j)$ on the grid is color coded depending on the value of the $R_{ij} = || x_{i} - x_{j}||$, where $1 \leq i, j \leq N$ and $|| . ||$ denotes the Euclidean norm.
 \textcolor{black}{
Let us consider the figure on the left of Fig. \ref{fig:RingStarone} (a). The $x$-state variable is color coded according to its value. This gives us an idea as to how the oscillators are evolving with time, are they synchronized with their neighboring elements or not? This can be seen if the oscillators have same value or color.   The leftmost plot illustrates the evolution of the network with time. The recurrence plot on the right, measures the Euclidean norm of the $i$ th oscillator node versus the $j$ th oscillator node. The color coding is done based on the Euclidean norm. If the norm between the $i$ th and $j$ th node is zero, then the nodes are synchronized. The shades represent the magnitude of the Euclidean norm of the $i$th oscillator versus the $j$th oscillator. When $i=j$, observe that the diagonal line is always blue denoting zero norm.}

When the ring coupling strength $\sigma$ is decreased to $-6$, the node gets synchronised. This is illustrated in Fig. \ref{fig:RingStarone} (b). Observe that the values of the $x$ state variable are all arranged horizontally. The SI value here is $0$ signifying synchronized state. This can also be confirmed from the spatiotemporal plot and the recurrence plot. 

The ring-star network of memristive Hindmarsh-Rose neuron system shows chimera state when $\sigma = 7, \mu =1$, see Fig. \ref{fig:RingStarone} (c). Observe that initial nodes and final nodes remain synchronised whereas nodes in the middle ($30\leq \rm Nodes \leq 60$ ) oscillate asynchronously. From such coexistence of synchronous and asynchronous states, it can confirmed as a chimera state. This can also be seen from the spatiotemporal plot on the left and the formation of the regular structures of colours other than blue in the recurrence plot. The SI value in this case is 0.7, confirming a chimera as $0< {\rm{SI}} < 1$.
\begin{figure*}[!htbp]
	\begin{center}
		\includegraphics[width=0.9\textwidth]{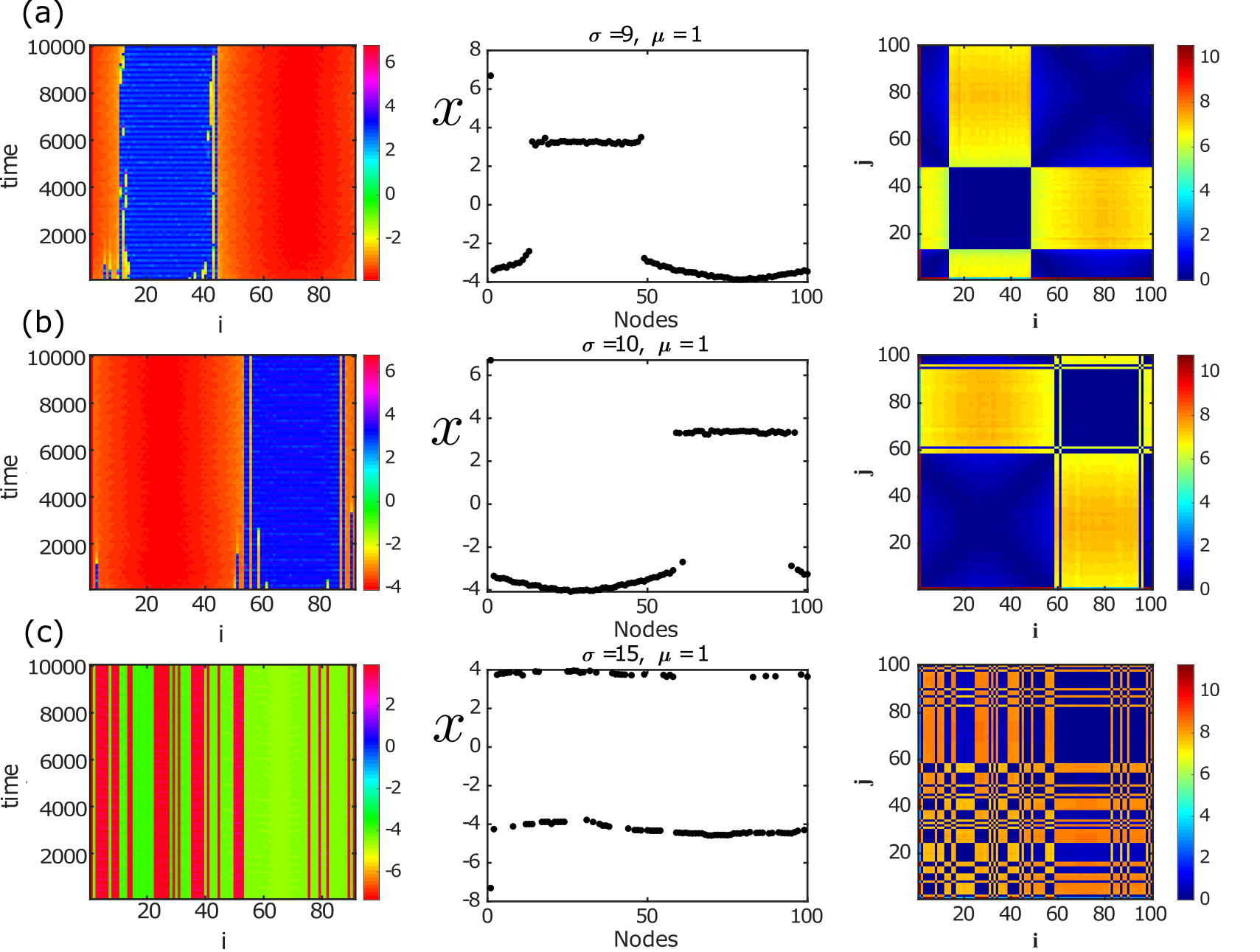}
	\end{center}
	\caption{Ring-Star network of memristive Hindmarsh Rose neuron model with $\sigma \neq 0, \mu \neq 0$. Random initial conditions are set with the coupling range of $P =70$. Number of nodes considered is $N =100$. Double well chimera in (a), another double well chimera in (b), two synchronized cluster state in (c) is shown.}
	\label{fig:RingStarTwo}
\end{figure*}
Interestingly, double-well chimera state is found in this system. We refer the reader to the author's previous work in \citep{ref51}, where double well chimera state were found in the ring-star network of Chua circuits. Double well chimera state is an important type of chimera state which traverses both the positive and negative values of $x$ state variable. It is interesting to observe in the case of memristive HR neuron system.

Such a double well chimera state is shown in Fig. \ref{fig:RingStarTwo} (a). Observe that in the middle plot, some nodes are in synchronous pattern in the positive range of $x$ and some in the negative range of $x$. Notice that the spatiotemporal pattern on the left, has alternating strips of both red (in negative region) and blue (in positive regions). The regular structures in the recurrence plot on the right also confirms this as a double well chimera state.The SI value in this case is 0.68 and denotes a chimera. Another such double well chimera state is shown in Fig. \ref{fig:RingStarTwo} (b). The SI value is 0.68, denoting a chimera state. The prevalence of the double-well chimera state was found to be robust with the variation of $\sigma$ till $\sigma < 15$. When $\sigma = 15, \mu =1$, the double well chimera state is destroyed  and formation of the two clustered state takes place as shown in Fig. \ref{fig:RingStarTwo} (c). The SI value is around 0.02 $\approx$ 0, signifying a cluster state. The middle plot shows the two almost synchronized clusters traversing both positive and negative values of $x$. The right most plot shows very tiny square like regular structures indicating the presence of clusters and the recurrence plot differs topologically form other patterns such as synchronous, chimera states. In the next section, we discuss about the topological patterns shown by the ring network.

\subsection{Ring network}
\label{sec:ringnet}
\begin{figure*}[!htbp]
	\begin{center}
		\includegraphics[width=0.9\textwidth]{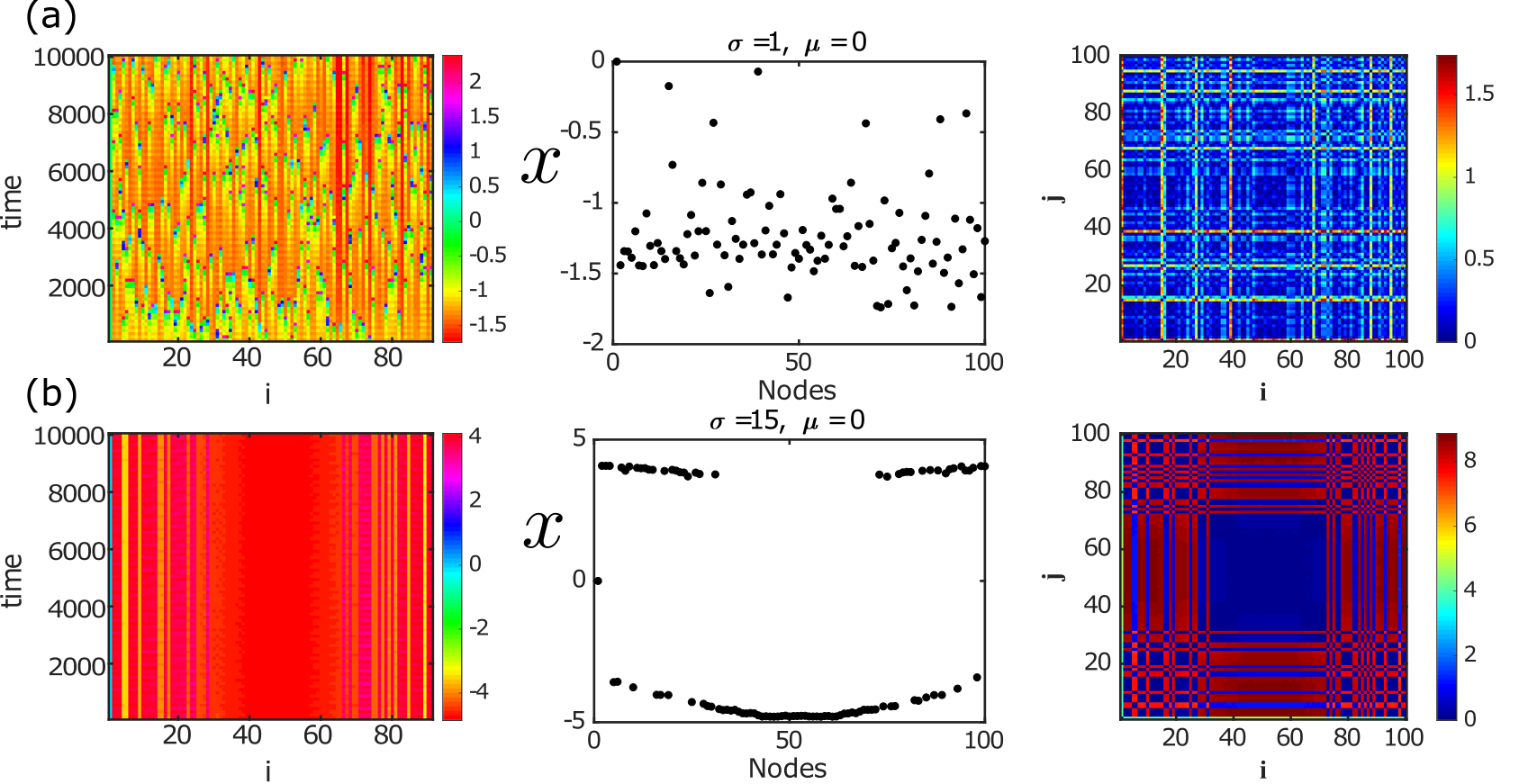}
	\end{center}
	\caption{Ring network of memristive Hindmarsh Rose neuron model with $\sigma \neq 0, \mu = 0$. Random initial conditions are set with the coupling range of $P =70$. Number of nodes considered is $N =100$. Asynchronous behavior in (a), double well chimera state in (b) is shown.}
	\label{fig:RingOne}
\end{figure*}

\begin{figure*}[!htbp]
	\begin{center}
		\includegraphics[width=0.9\textwidth]{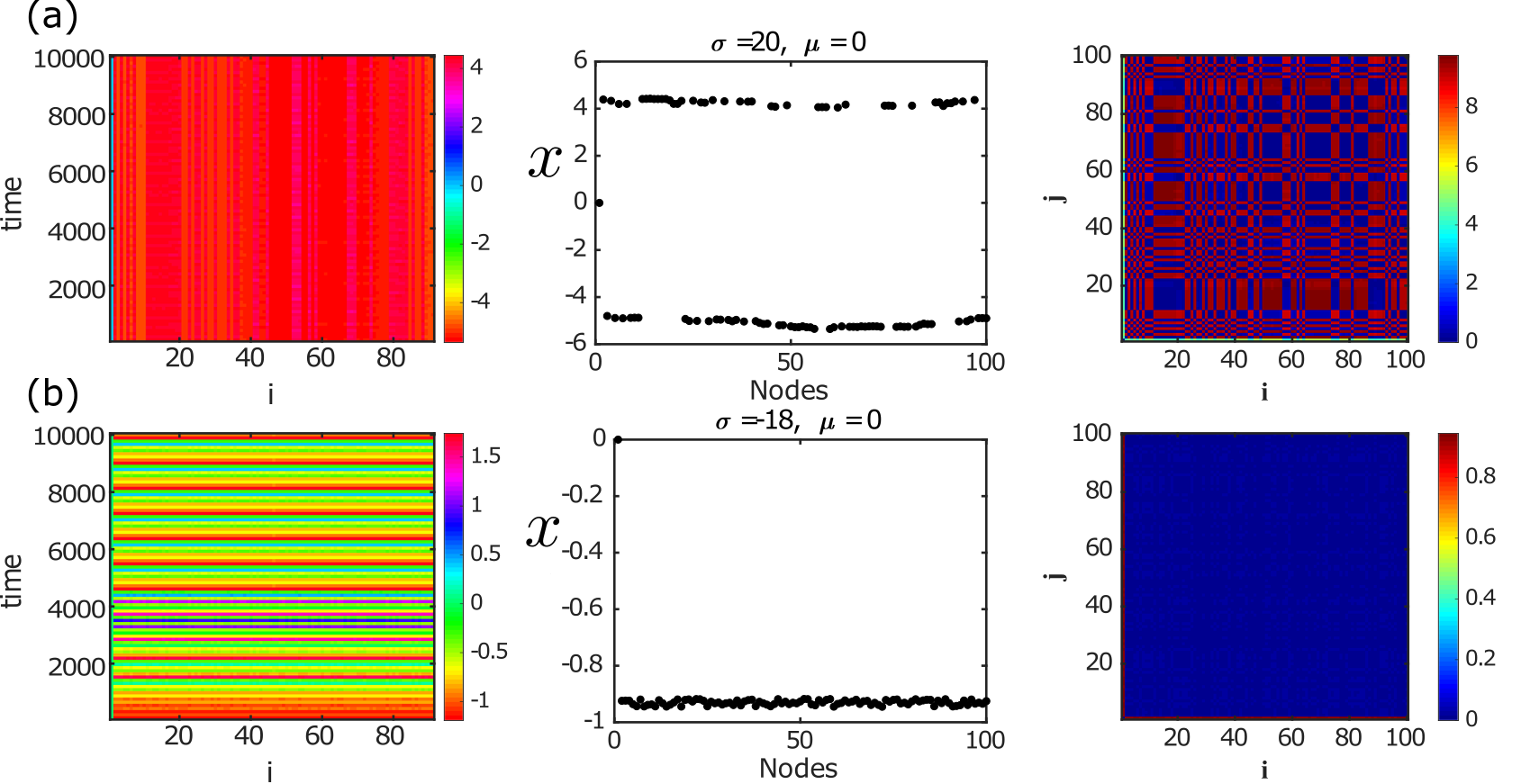}
	\end{center}
	\caption{Ring network of memristive Hindmarsh Rose neuron model with $\sigma \neq 0, \mu = 0$. Random initial conditions are set with the coupling range of $P =70$. Number of nodes considered is $N =100$. Two cluster synchronized state behavior in (a), synchronized state in (b) is shown.}
	\label{fig:RingTwo}
\end{figure*}

In this section, we address various spatiotemporal patterns in the ring network of memristive Hindmarsh Rose neuron system by setting $\sigma \neq 0, \mu  =0$. In Fig. \ref{fig:RingOne} (a), we showcase the asynchronous behavior of the end nodes of the ring network. This can be confirmed from the spatiotemporal plot and the recurrence plot. The SI value is 1 denoting an asynchronous state. In Fig. \ref{fig:RingOne} (b), we showcase a double well chimera state. This can be confirmed by the regular structures in the recurrence plot on the right and the spatiotemporal plot on the left. The SI value is 0.44, signifying  a chimera state.
Moreover, cluster states are also possible in ring network, see Fig. \ref{fig:RingTwo} (a). The SI value is 0.02, signifying synchronous state/ cluster state. This can be shown by the presence of small square structures in the recurrence plots. A single cluster synchronization state is shown in Fig. \ref{fig:RingTwo} (b).The SI value is 0, denoting a synchronized state. 

\subsection{Star network}
\label{sec:starnet}
In this section, we explore the spatiotemporal patterns exhbited by the star network of memristive Hindmarsh Rose neuron model. Star networks is useful in many engineering systems, network hub system. Study on the synchronization aspect of star connected Chua oscillator were carried out in \citep{ref51}.  Unlike previous cases of ring-star and ring network, chimera state seems to be absent in the case of star networks. The presence of sole central node drives more information to the end nodes of the network and hence chances of full synchronization is much more common in star networks. 
When $\mu = -0.5$, the star network enters the regime of asynchronization. The SI value is 1, signifying asynchronization. Increasing the star coupling strength $\mu$ to $1$, we observe full synchronization in the system.The SI value is 0, signifying synchronization.
\begin{figure*}[!htbp]
	\begin{center}
		\includegraphics[width=0.9\textwidth]{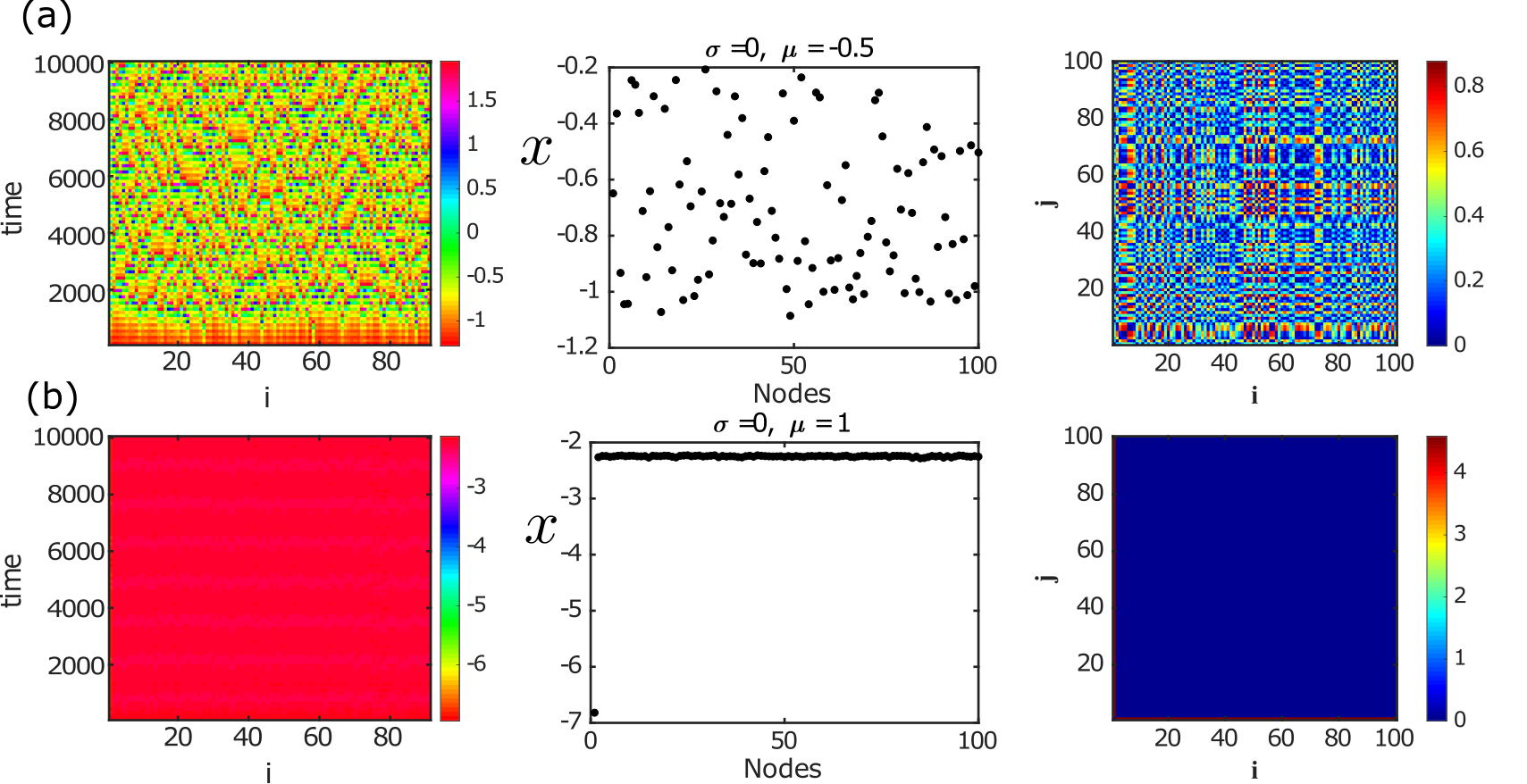}
	\end{center}
	\caption{Star network of memristive Hindmarsh Rose neuron model with $\sigma = 0, \mu \neq 0$. Random initial conditions are set with the coupling range of $P =70$. Number of nodes considered is $N =100$. Asynchronous behavior in (a), synchronized state in (b) is shown.}
	\label{fig:StarOne}
\end{figure*}

\subsection{Variation of the Strength of Incoherence with respect to coupling strength}
\begin{figure*}[!htbp]
	\begin{center}
	\includegraphics[width=0.9\textwidth]{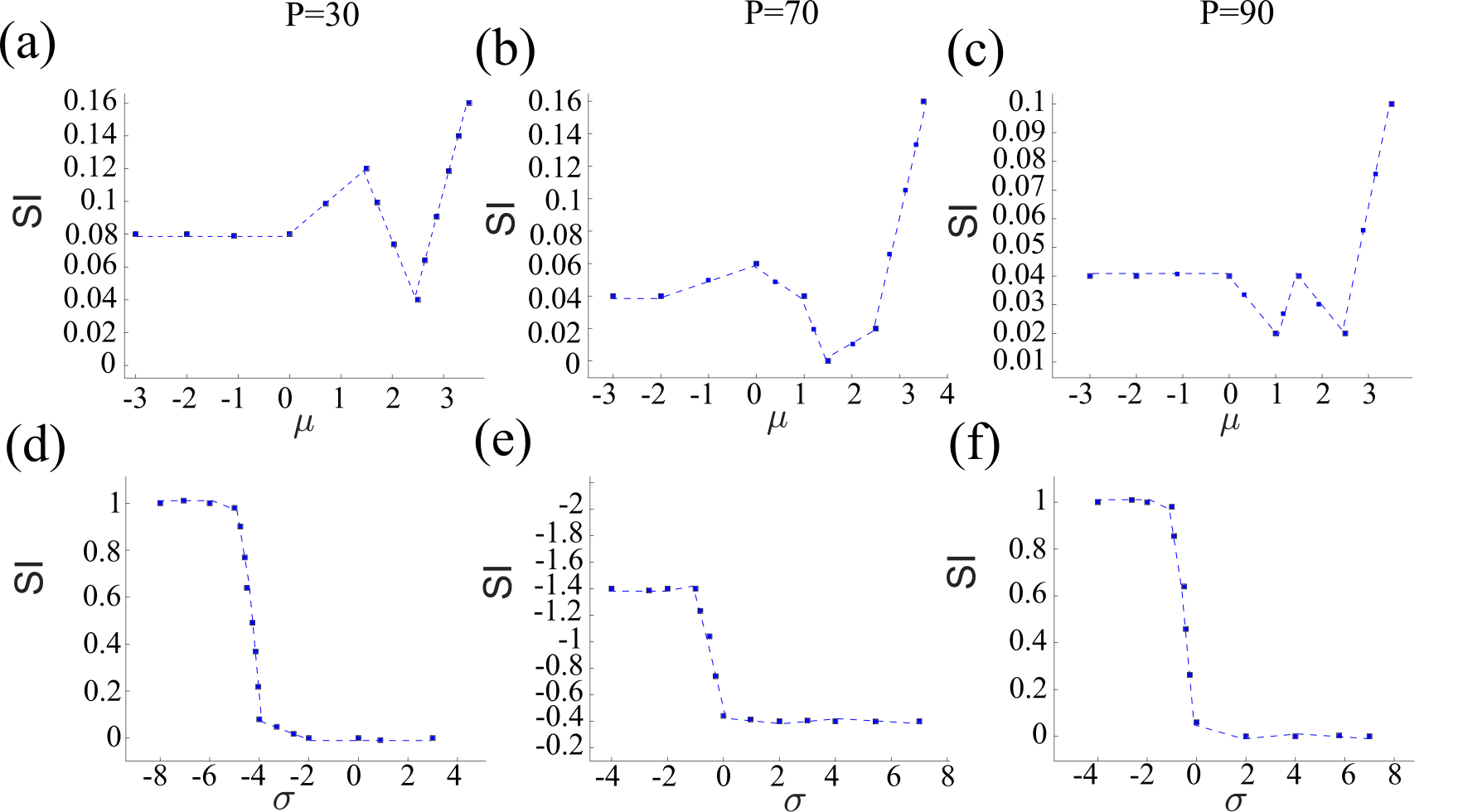}
		\end{center}
 	\caption{Variation of the strength of incoherence (SI) with respect to the star coupling strength ($\mu$) in panels (a), (b), (c) for various coupling ranges $P = 30, 70,$ and $90$ respectively. Similarly, variation of the strength of incoherence (SI) with respect to the ring coupling strength $\sigma$ for various values of coupling ranges $P=30, 70,$ and $90$ respectively. The  network size is $N=100$.}
	\label{fig:SI_Variation_CouplingStrength}
\end{figure*}

Here we observe the variation of the strength of incoherence (SI) with respect to the ring coupling strength ($\sigma$), star coupling strength ($\mu$). In Fig. \ref{fig:SI_Variation_CouplingStrength} (a), (b), and (c), we have considered the variation of the strength of incoherence with the variation of the star coupling strength $\mu$ for three different values of coupling range $P = 30, 70, {\rm{and}} \, 90$. In Fig. \ref{fig:SI_Variation_CouplingStrength} (a), with negative $\mu$, the SI value is almost same and then increases as $\mu$ becomes positive and then follows an increasing trend as $\mu$ is increased. The behavior is robust with the change in the coupling range $P$ as evident from Fig. \ref{fig:SI_Variation_CouplingStrength} (b), (c). 

In Fig. \ref{fig:SI_Variation_CouplingStrength} (d), (e), (f), we have considered the variation of the strength of incoherence with the variation of ring coupling strength $\sigma$ for three different values of coupling range $P = 30, 70, {\rm{and}} \, 90$. As can be seen in Fig. \ref{fig:SI_Variation_CouplingStrength} (d), SI is $1$ for negative values of $\sigma$ and starts to decrease for positive $\sigma$ and reaches to zero. So variation of $\sigma$ from negative to positive value, we see a variation from asynchronous to synchronous state or cluster state. The behavior is robust for different other values of coupling ranges in Fig. \ref{fig:SI_Variation_CouplingStrength} (e), and (f).  

	\section{Conclusion}
	
In this manuscript, we have considered a memristive version of the Hindmarsh-Rose neuron model. We found the proposed model was able to exhibit a reverse period doubling route to chaos, as well as phenomena of interior and exterior crises. Three different networks (ring-star, ring, and star) networks of memristive Hindmarsh Rose neuron models were explored. Chimera states, including double well chimera states, were found in the ring-star and ring network, which shows that the memristive Hindmarsh-Rose neuron model is a promising neuron model to be explored further in the future. Many future directions emerge from this study. Study of lattice \cite{shepelev2020role, shepelev2021spatiotemporal}, multilayer networks \cite{shepelev2021synchronization} of the neuron model can be explored. Emergence of spiral waves can be studied in the latter networks and a proper quantification can be carried similar to the methods used in \cite{shepelev2020quantifying}. The basin of attraction of double-well chimera state, synchronous and asynchronous state, can be explored in the future. Does the system exhibit anti-phase synchronization \cite{shepelev2021repulsive} is a topic that can be thought of. Does a discretized version of the proposed model in the present paper show extra qualitative dynamics is a future direction that can be looked upon in a similar spirit in \cite{muni2022dynamical}. Recently extreme multistability was found in memrisitve Hindmarsh-Rose model in \cite{njitacke2022coexistence}, can this model also exhibit extreme multistability?. A deep investigation of the global dynamics of the memristive Hindmarsh-Rose neuron proposed in this work will be carried out.

\section*{Acknowledgments}
This work is partially funded by the Polish National Science Center under the Grant OPUS $14 No.2017/27/B/ST8/01330$.

\bibliographystyle{ieeetr}
\bibliography{Projet-1}

\begin{thebibliography}{65}
\expandafter\ifx\csname natexlab\endcsname\relax\def\natexlab#1{#1}\fi
\expandafter\ifx\csname bibnamefont\endcsname\relax
  \def\bibnamefont#1{#1}\fi
\expandafter\ifx\csname bibfnamefont\endcsname\relax
  \def\bibfnamefont#1{#1}\fi
\expandafter\ifx\csname citenamefont\endcsname\relax
  \def\citenamefont#1{#1}\fi
\expandafter\ifx\csname url\endcsname\relax
  \def\url#1{\texttt{#1}}\fi
\expandafter\ifx\csname urlprefix\endcsname\relax\def\urlprefix{URL }\fi
\providecommand{\bibinfo}[2]{#2}
\providecommand{\eprint}[2][]{\url{#2}}

\bibitem[{\citenamefont{Lin et~al.}(2021)\citenamefont{Lin, Wang, Deng, Xu,
  Deng, and Zhou}}]{ref1}
\bibinfo{author}{\bibfnamefont{H.}~\bibnamefont{Lin}},
  \bibinfo{author}{\bibfnamefont{C.}~\bibnamefont{Wang}},
  \bibinfo{author}{\bibfnamefont{Q.}~\bibnamefont{Deng}},
  \bibinfo{author}{\bibfnamefont{C.}~\bibnamefont{Xu}},
  \bibinfo{author}{\bibfnamefont{Z.}~\bibnamefont{Deng}}, \bibnamefont{and}
  \bibinfo{author}{\bibfnamefont{C.}~\bibnamefont{Zhou}},
  \bibinfo{journal}{Nonlinear Dynamics} \textbf{\bibinfo{volume}{106}},
  \bibinfo{pages}{959} (\bibinfo{year}{2021}).

\bibitem[{\citenamefont{Zhang et~al.}(2018)\citenamefont{Zhang, Wang,
  Alzahrani, Wu, and An}}]{ref2}
\bibinfo{author}{\bibfnamefont{G.}~\bibnamefont{Zhang}},
  \bibinfo{author}{\bibfnamefont{C.}~\bibnamefont{Wang}},
  \bibinfo{author}{\bibfnamefont{F.}~\bibnamefont{Alzahrani}},
  \bibinfo{author}{\bibfnamefont{F.}~\bibnamefont{Wu}}, \bibnamefont{and}
  \bibinfo{author}{\bibfnamefont{X.}~\bibnamefont{An}},
  \bibinfo{journal}{Chaos, Solitons \& Fractals}
  \textbf{\bibinfo{volume}{108}}, \bibinfo{pages}{15} (\bibinfo{year}{2018}).

\bibitem[{\citenamefont{Njitacke
  et~al.}(2021{\natexlab{a}})\citenamefont{Njitacke, Isaac, Nestor, and
  Kengne}}]{ref3}
\bibinfo{author}{\bibfnamefont{Z.~T.} \bibnamefont{Njitacke}},
  \bibinfo{author}{\bibfnamefont{S.~D.} \bibnamefont{Isaac}},
  \bibinfo{author}{\bibfnamefont{T.}~\bibnamefont{Nestor}}, \bibnamefont{and}
  \bibinfo{author}{\bibfnamefont{J.}~\bibnamefont{Kengne}},
  \bibinfo{journal}{Neural Computing and Applications}
  \textbf{\bibinfo{volume}{33}}, \bibinfo{pages}{6733}
  (\bibinfo{year}{2021}{\natexlab{a}}).

\bibitem[{\citenamefont{Tabekoueng~Njitacke
  et~al.}(2020{\natexlab{a}})\citenamefont{Tabekoueng~Njitacke, Kengne, and
  Fotsin}}]{ref4}
\bibinfo{author}{\bibfnamefont{Z.}~\bibnamefont{Tabekoueng~Njitacke}},
  \bibinfo{author}{\bibfnamefont{J.}~\bibnamefont{Kengne}}, \bibnamefont{and}
  \bibinfo{author}{\bibfnamefont{H.~B.} \bibnamefont{Fotsin}},
  \bibinfo{journal}{Circuits, Systems, and Signal Processing}
  \textbf{\bibinfo{volume}{39}}, \bibinfo{pages}{3424}
  (\bibinfo{year}{2020}{\natexlab{a}}).

\bibitem[{\citenamefont{Doubla~Isaac et~al.}(2020)\citenamefont{Doubla~Isaac,
  Njitacke, and Kengne}}]{ref5}
\bibinfo{author}{\bibfnamefont{S.}~\bibnamefont{Doubla~Isaac}},
  \bibinfo{author}{\bibfnamefont{Z.~T.} \bibnamefont{Njitacke}},
  \bibnamefont{and} \bibinfo{author}{\bibfnamefont{J.}~\bibnamefont{Kengne}},
  \bibinfo{journal}{International Journal of Bifurcation and Chaos}
  \textbf{\bibinfo{volume}{30}}, \bibinfo{pages}{2050159}
  (\bibinfo{year}{2020}).

\bibitem[{\citenamefont{Tabekoueng~Njitacke
  et~al.}(2020{\natexlab{b}})\citenamefont{Tabekoueng~Njitacke, Laura~Matze,
  Fouodji~Tsotsop, and Kengne}}]{ref6}
\bibinfo{author}{\bibfnamefont{Z.}~\bibnamefont{Tabekoueng~Njitacke}},
  \bibinfo{author}{\bibfnamefont{C.}~\bibnamefont{Laura~Matze}},
  \bibinfo{author}{\bibfnamefont{M.}~\bibnamefont{Fouodji~Tsotsop}},
  \bibnamefont{and} \bibinfo{author}{\bibfnamefont{J.}~\bibnamefont{Kengne}},
  \bibinfo{journal}{Neural Processing Letters} \textbf{\bibinfo{volume}{52}},
  \bibinfo{pages}{267} (\bibinfo{year}{2020}{\natexlab{b}}).

\bibitem[{\citenamefont{Hodgkin and Huxley}(1990)}]{ref7}
\bibinfo{author}{\bibfnamefont{A.}~\bibnamefont{Hodgkin}} \bibnamefont{and}
  \bibinfo{author}{\bibfnamefont{A.}~\bibnamefont{Huxley}},
  \bibinfo{journal}{Bulletin of mathematical biology}
  \textbf{\bibinfo{volume}{52}}, \bibinfo{pages}{25} (\bibinfo{year}{1990}).

\bibitem[{\citenamefont{Hindmarsh and Rose}(1982)}]{ref12}
\bibinfo{author}{\bibfnamefont{J.}~\bibnamefont{Hindmarsh}} \bibnamefont{and}
  \bibinfo{author}{\bibfnamefont{R.}~\bibnamefont{Rose}},
  \bibinfo{journal}{Nature} \textbf{\bibinfo{volume}{296}},
  \bibinfo{pages}{162} (\bibinfo{year}{1982}).

\bibitem[{\citenamefont{Hindmarsh and Rose}(1984)}]{ref13}
\bibinfo{author}{\bibfnamefont{J.~L.} \bibnamefont{Hindmarsh}}
  \bibnamefont{and} \bibinfo{author}{\bibfnamefont{R.}~\bibnamefont{Rose}},
  \bibinfo{journal}{Proceedings of the Royal society of London. Series B.
  Biological sciences} \textbf{\bibinfo{volume}{221}}, \bibinfo{pages}{87}
  (\bibinfo{year}{1984}).

\bibitem[{\citenamefont{Izhikevich and FitzHugh}(2006)}]{ref10}
\bibinfo{author}{\bibfnamefont{E.~M.} \bibnamefont{Izhikevich}}
  \bibnamefont{and} \bibinfo{author}{\bibfnamefont{R.}~\bibnamefont{FitzHugh}},
  \bibinfo{journal}{Scholarpedia} \textbf{\bibinfo{volume}{1}},
  \bibinfo{pages}{1349} (\bibinfo{year}{2006}).

\bibitem[{\citenamefont{Korkmaz and Sivga}(2022)}]{korkmaz4fpga}
\bibinfo{author}{\bibfnamefont{N.}~\bibnamefont{Korkmaz}} \bibnamefont{and}
  \bibinfo{author}{\bibfnamefont{B.}~\bibnamefont{Sivga}},
  \bibinfo{journal}{Chaos Theory and Applications}
  \textbf{\bibinfo{volume}{4}}, \bibinfo{pages}{88} (\bibinfo{year}{2022}).

\bibitem[{\citenamefont{Tsumoto et~al.}(2006)\citenamefont{Tsumoto, Kitajima,
  Yoshinaga, Aihara, and Kawakami}}]{ref11}
\bibinfo{author}{\bibfnamefont{K.}~\bibnamefont{Tsumoto}},
  \bibinfo{author}{\bibfnamefont{H.}~\bibnamefont{Kitajima}},
  \bibinfo{author}{\bibfnamefont{T.}~\bibnamefont{Yoshinaga}},
  \bibinfo{author}{\bibfnamefont{K.}~\bibnamefont{Aihara}}, \bibnamefont{and}
  \bibinfo{author}{\bibfnamefont{H.}~\bibnamefont{Kawakami}},
  \bibinfo{journal}{Neurocomputing} \textbf{\bibinfo{volume}{69}},
  \bibinfo{pages}{293} (\bibinfo{year}{2006}).

\bibitem[{\citenamefont{Chay}(1985)}]{ref8}
\bibinfo{author}{\bibfnamefont{T.~R.} \bibnamefont{Chay}},
  \bibinfo{journal}{Physica D: Nonlinear Phenomena}
  \textbf{\bibinfo{volume}{16}}, \bibinfo{pages}{233} (\bibinfo{year}{1985}).

\bibitem[{\citenamefont{Izhikevich}(2003)}]{ref9}
\bibinfo{author}{\bibfnamefont{E.~M.} \bibnamefont{Izhikevich}},
  \bibinfo{journal}{IEEE Transactions on neural networks}
  \textbf{\bibinfo{volume}{14}}, \bibinfo{pages}{1569} (\bibinfo{year}{2003}).

\bibitem[{\citenamefont{Xu et~al.}(2021)\citenamefont{Xu, Liu, Feng, Bao, Wu,
  and Bao}}]{ref14}
\bibinfo{author}{\bibfnamefont{Q.}~\bibnamefont{Xu}},
  \bibinfo{author}{\bibfnamefont{T.}~\bibnamefont{Liu}},
  \bibinfo{author}{\bibfnamefont{C.-T.} \bibnamefont{Feng}},
  \bibinfo{author}{\bibfnamefont{H.}~\bibnamefont{Bao}},
  \bibinfo{author}{\bibfnamefont{H.-G.} \bibnamefont{Wu}}, \bibnamefont{and}
  \bibinfo{author}{\bibfnamefont{B.-C.} \bibnamefont{Bao}},
  \bibinfo{journal}{Chinese Physics B} \textbf{\bibinfo{volume}{30}},
  \bibinfo{pages}{128702} (\bibinfo{year}{2021}).

\bibitem[{\citenamefont{Liu et~al.}(2019)\citenamefont{Liu, Wang, Zhang, and
  Zhang}}]{ref18}
\bibinfo{author}{\bibfnamefont{Z.}~\bibnamefont{Liu}},
  \bibinfo{author}{\bibfnamefont{C.}~\bibnamefont{Wang}},
  \bibinfo{author}{\bibfnamefont{G.}~\bibnamefont{Zhang}}, \bibnamefont{and}
  \bibinfo{author}{\bibfnamefont{Y.}~\bibnamefont{Zhang}},
  \bibinfo{journal}{International Journal of Modern Physics B}
  \textbf{\bibinfo{volume}{33}}, \bibinfo{pages}{1950170}
  (\bibinfo{year}{2019}).

\bibitem[{\citenamefont{Zhang et~al.}(2020{\natexlab{a}})\citenamefont{Zhang,
  Wang, Tang, Ma, and Ren}}]{ref19}
\bibinfo{author}{\bibfnamefont{Y.}~\bibnamefont{Zhang}},
  \bibinfo{author}{\bibfnamefont{C.}~\bibnamefont{Wang}},
  \bibinfo{author}{\bibfnamefont{J.}~\bibnamefont{Tang}},
  \bibinfo{author}{\bibfnamefont{J.}~\bibnamefont{Ma}}, \bibnamefont{and}
  \bibinfo{author}{\bibfnamefont{G.}~\bibnamefont{Ren}},
  \bibinfo{journal}{Science China Technological Sciences}
  \textbf{\bibinfo{volume}{63}}, \bibinfo{pages}{2328}
  (\bibinfo{year}{2020}{\natexlab{a}}).

\bibitem[{\citenamefont{Li}(2021)}]{ref20}
\bibinfo{author}{\bibfnamefont{Y.}~\bibnamefont{Li}},
  \bibinfo{journal}{Physical Review Research} \textbf{\bibinfo{volume}{3}},
  \bibinfo{pages}{023146} (\bibinfo{year}{2021}).

\bibitem[{\citenamefont{Shaffer et~al.}(2016)\citenamefont{Shaffer, Harris,
  Follmann, and Rosa~Jr}}]{ref16}
\bibinfo{author}{\bibfnamefont{A.}~\bibnamefont{Shaffer}},
  \bibinfo{author}{\bibfnamefont{A.~L.} \bibnamefont{Harris}},
  \bibinfo{author}{\bibfnamefont{R.}~\bibnamefont{Follmann}}, \bibnamefont{and}
  \bibinfo{author}{\bibfnamefont{E.}~\bibnamefont{Rosa~Jr}},
  \bibinfo{journal}{Physical Review E} \textbf{\bibinfo{volume}{94}},
  \bibinfo{pages}{042301} (\bibinfo{year}{2016}).

\bibitem[{\citenamefont{Zhou et~al.}(2021{\natexlab{a}})\citenamefont{Zhou,
  Jiang, Xu, Xu, Zhou, and Yuan}}]{ref17}
\bibinfo{author}{\bibfnamefont{J.-F.} \bibnamefont{Zhou}},
  \bibinfo{author}{\bibfnamefont{E.-H.} \bibnamefont{Jiang}},
  \bibinfo{author}{\bibfnamefont{B.-L.} \bibnamefont{Xu}},
  \bibinfo{author}{\bibfnamefont{K.}~\bibnamefont{Xu}},
  \bibinfo{author}{\bibfnamefont{C.}~\bibnamefont{Zhou}}, \bibnamefont{and}
  \bibinfo{author}{\bibfnamefont{W.-J.} \bibnamefont{Yuan}},
  \bibinfo{journal}{Physical Review E} \textbf{\bibinfo{volume}{104}},
  \bibinfo{pages}{054407} (\bibinfo{year}{2021}{\natexlab{a}}).

\bibitem[{\citenamefont{Buri{\'c} et~al.}(2008)\citenamefont{Buri{\'c},
  Todorovi{\'c}, and Vasovi{\'c}}}]{ref15}
\bibinfo{author}{\bibfnamefont{N.}~\bibnamefont{Buri{\'c}}},
  \bibinfo{author}{\bibfnamefont{K.}~\bibnamefont{Todorovi{\'c}}},
  \bibnamefont{and}
  \bibinfo{author}{\bibfnamefont{N.}~\bibnamefont{Vasovi{\'c}}},
  \bibinfo{journal}{Physical Review E} \textbf{\bibinfo{volume}{78}},
  \bibinfo{pages}{036211} (\bibinfo{year}{2008}).

\bibitem[{\citenamefont{Bao et~al.}(2018)\citenamefont{Bao, Hu, Xu, Bao, Wu,
  and Chen}}]{ref21}
\bibinfo{author}{\bibfnamefont{B.}~\bibnamefont{Bao}},
  \bibinfo{author}{\bibfnamefont{A.}~\bibnamefont{Hu}},
  \bibinfo{author}{\bibfnamefont{Q.}~\bibnamefont{Xu}},
  \bibinfo{author}{\bibfnamefont{H.}~\bibnamefont{Bao}},
  \bibinfo{author}{\bibfnamefont{H.}~\bibnamefont{Wu}}, \bibnamefont{and}
  \bibinfo{author}{\bibfnamefont{M.}~\bibnamefont{Chen}},
  \bibinfo{journal}{Nonlinear Dynamics} \textbf{\bibinfo{volume}{92}},
  \bibinfo{pages}{1695} (\bibinfo{year}{2018}).

\bibitem[{\citenamefont{Bao et~al.}(2019)\citenamefont{Bao, Hu, Liu, and
  Bao}}]{ref22}
\bibinfo{author}{\bibfnamefont{H.}~\bibnamefont{Bao}},
  \bibinfo{author}{\bibfnamefont{A.}~\bibnamefont{Hu}},
  \bibinfo{author}{\bibfnamefont{W.}~\bibnamefont{Liu}}, \bibnamefont{and}
  \bibinfo{author}{\bibfnamefont{B.}~\bibnamefont{Bao}}, \bibinfo{journal}{IEEE
  transactions on neural networks and learning systems}
  \textbf{\bibinfo{volume}{31}}, \bibinfo{pages}{502} (\bibinfo{year}{2019}).

\bibitem[{\citenamefont{Hou et~al.}(2021)\citenamefont{Hou, Ma, Zhan, Yang, and
  Jia}}]{ref23}
\bibinfo{author}{\bibfnamefont{Z.}~\bibnamefont{Hou}},
  \bibinfo{author}{\bibfnamefont{J.}~\bibnamefont{Ma}},
  \bibinfo{author}{\bibfnamefont{X.}~\bibnamefont{Zhan}},
  \bibinfo{author}{\bibfnamefont{L.}~\bibnamefont{Yang}}, \bibnamefont{and}
  \bibinfo{author}{\bibfnamefont{Y.}~\bibnamefont{Jia}},
  \bibinfo{journal}{Chaos, Solitons \& Fractals}
  \textbf{\bibinfo{volume}{142}}, \bibinfo{pages}{110522}
  (\bibinfo{year}{2021}).

\bibitem[{\citenamefont{Liu et~al.}(2020)\citenamefont{Liu, Xu, Ma, Alzahrani,
  and Hobiny}}]{ref24}
\bibinfo{author}{\bibfnamefont{Y.}~\bibnamefont{Liu}},
  \bibinfo{author}{\bibfnamefont{W.-j.} \bibnamefont{Xu}},
  \bibinfo{author}{\bibfnamefont{J.}~\bibnamefont{Ma}},
  \bibinfo{author}{\bibfnamefont{F.}~\bibnamefont{Alzahrani}},
  \bibnamefont{and} \bibinfo{author}{\bibfnamefont{A.}~\bibnamefont{Hobiny}},
  \bibinfo{journal}{Frontiers of Information Technology \& Electronic
  Engineering} \textbf{\bibinfo{volume}{21}}, \bibinfo{pages}{1387}
  (\bibinfo{year}{2020}).

\bibitem[{\citenamefont{Zhang et~al.}(2020{\natexlab{b}})\citenamefont{Zhang,
  Guo, Wu, and Ma}}]{ref25}
\bibinfo{author}{\bibfnamefont{G.}~\bibnamefont{Zhang}},
  \bibinfo{author}{\bibfnamefont{D.}~\bibnamefont{Guo}},
  \bibinfo{author}{\bibfnamefont{F.}~\bibnamefont{Wu}}, \bibnamefont{and}
  \bibinfo{author}{\bibfnamefont{J.}~\bibnamefont{Ma}},
  \bibinfo{journal}{Neurocomputing} \textbf{\bibinfo{volume}{379}},
  \bibinfo{pages}{296} (\bibinfo{year}{2020}{\natexlab{b}}).

\bibitem[{\citenamefont{Zhou et~al.}(2021{\natexlab{b}})\citenamefont{Zhou,
  Yao, Ma, and Zhu}}]{ref26}
\bibinfo{author}{\bibfnamefont{P.}~\bibnamefont{Zhou}},
  \bibinfo{author}{\bibfnamefont{Z.}~\bibnamefont{Yao}},
  \bibinfo{author}{\bibfnamefont{J.}~\bibnamefont{Ma}}, \bibnamefont{and}
  \bibinfo{author}{\bibfnamefont{Z.}~\bibnamefont{Zhu}},
  \bibinfo{journal}{Chaos, Solitons \& Fractals}
  \textbf{\bibinfo{volume}{145}}, \bibinfo{pages}{110751}
  (\bibinfo{year}{2021}{\natexlab{b}}).

\bibitem[{\citenamefont{Cai et~al.}(2021)\citenamefont{Cai, Bao, Xu, Hua, and
  Bao}}]{ref27}
\bibinfo{author}{\bibfnamefont{J.}~\bibnamefont{Cai}},
  \bibinfo{author}{\bibfnamefont{H.}~\bibnamefont{Bao}},
  \bibinfo{author}{\bibfnamefont{Q.}~\bibnamefont{Xu}},
  \bibinfo{author}{\bibfnamefont{Z.}~\bibnamefont{Hua}}, \bibnamefont{and}
  \bibinfo{author}{\bibfnamefont{B.}~\bibnamefont{Bao}},
  \bibinfo{journal}{Nonlinear Dynamics} \textbf{\bibinfo{volume}{104}},
  \bibinfo{pages}{4379} (\bibinfo{year}{2021}).

\bibitem[{\citenamefont{Li et~al.}(2021{\natexlab{a}})\citenamefont{Li, Bao,
  Li, Ma, Hua, and Bao}}]{ref28}
\bibinfo{author}{\bibfnamefont{K.}~\bibnamefont{Li}},
  \bibinfo{author}{\bibfnamefont{H.}~\bibnamefont{Bao}},
  \bibinfo{author}{\bibfnamefont{H.}~\bibnamefont{Li}},
  \bibinfo{author}{\bibfnamefont{J.}~\bibnamefont{Ma}},
  \bibinfo{author}{\bibfnamefont{Z.}~\bibnamefont{Hua}}, \bibnamefont{and}
  \bibinfo{author}{\bibfnamefont{B.}~\bibnamefont{Bao}}, \bibinfo{journal}{IEEE
  Transactions on Industrial Informatics} \textbf{\bibinfo{volume}{18}},
  \bibinfo{pages}{1726} (\bibinfo{year}{2021}{\natexlab{a}}).

\bibitem[{\citenamefont{Njitacke
  et~al.}(2022{\natexlab{a}})\citenamefont{Njitacke, Awrejcewicz, Ramakrishnan,
  Rajagopal, and Kengne}}]{ref29}
\bibinfo{author}{\bibfnamefont{Z.~T.} \bibnamefont{Njitacke}},
  \bibinfo{author}{\bibfnamefont{J.}~\bibnamefont{Awrejcewicz}},
  \bibinfo{author}{\bibfnamefont{B.}~\bibnamefont{Ramakrishnan}},
  \bibinfo{author}{\bibfnamefont{K.}~\bibnamefont{Rajagopal}},
  \bibnamefont{and} \bibinfo{author}{\bibfnamefont{J.}~\bibnamefont{Kengne}},
  \bibinfo{journal}{Nonlinear Dynamics} \textbf{\bibinfo{volume}{107}},
  \bibinfo{pages}{2867} (\bibinfo{year}{2022}{\natexlab{a}}).

\bibitem[{\citenamefont{Njitacke et~al.}(2020)\citenamefont{Njitacke, Doubla,
  Mabekou, and Kengne}}]{ref30}
\bibinfo{author}{\bibfnamefont{Z.~T.} \bibnamefont{Njitacke}},
  \bibinfo{author}{\bibfnamefont{I.~S.} \bibnamefont{Doubla}},
  \bibinfo{author}{\bibfnamefont{S.}~\bibnamefont{Mabekou}}, \bibnamefont{and}
  \bibinfo{author}{\bibfnamefont{J.}~\bibnamefont{Kengne}},
  \bibinfo{journal}{Chaos, Solitons \& Fractals}
  \textbf{\bibinfo{volume}{137}}, \bibinfo{pages}{109785}
  (\bibinfo{year}{2020}).

\bibitem[{\citenamefont{Njitacke
  et~al.}(2021{\natexlab{b}})\citenamefont{Njitacke, Koumetio, Ramakrishnan,
  Leutcho, Fozin, Tsafack, Rajagopal, and Kengne}}]{ref31}
\bibinfo{author}{\bibfnamefont{Z.~T.} \bibnamefont{Njitacke}},
  \bibinfo{author}{\bibfnamefont{B.~N.} \bibnamefont{Koumetio}},
  \bibinfo{author}{\bibfnamefont{B.}~\bibnamefont{Ramakrishnan}},
  \bibinfo{author}{\bibfnamefont{G.~D.} \bibnamefont{Leutcho}},
  \bibinfo{author}{\bibfnamefont{T.~F.} \bibnamefont{Fozin}},
  \bibinfo{author}{\bibfnamefont{N.}~\bibnamefont{Tsafack}},
  \bibinfo{author}{\bibfnamefont{K.}~\bibnamefont{Rajagopal}},
  \bibnamefont{and} \bibinfo{author}{\bibfnamefont{J.}~\bibnamefont{Kengne}},
  \bibinfo{journal}{Cognitive Neurodynamics} pp. \bibinfo{pages}{1--18}
  (\bibinfo{year}{2021}{\natexlab{b}}).

\bibitem[{\citenamefont{Njitacke
  et~al.}(2021{\natexlab{c}})\citenamefont{Njitacke, Tsafack, Ramakrishnan,
  Rajagopal, Kengne, and Awrejcewicz}}]{ref32}
\bibinfo{author}{\bibfnamefont{Z.~T.} \bibnamefont{Njitacke}},
  \bibinfo{author}{\bibfnamefont{N.}~\bibnamefont{Tsafack}},
  \bibinfo{author}{\bibfnamefont{B.}~\bibnamefont{Ramakrishnan}},
  \bibinfo{author}{\bibfnamefont{K.}~\bibnamefont{Rajagopal}},
  \bibinfo{author}{\bibfnamefont{J.}~\bibnamefont{Kengne}}, \bibnamefont{and}
  \bibinfo{author}{\bibfnamefont{J.}~\bibnamefont{Awrejcewicz}},
  \bibinfo{journal}{Chaos, Solitons \& Fractals}
  \textbf{\bibinfo{volume}{153}}, \bibinfo{pages}{111577}
  (\bibinfo{year}{2021}{\natexlab{c}}).

\bibitem[{\citenamefont{Tabekoueng~Njitacke
  et~al.}(2020{\natexlab{c}})\citenamefont{Tabekoueng~Njitacke, Sami~Doubla,
  Kengne, and Cheukem}}]{ref33}
\bibinfo{author}{\bibfnamefont{Z.}~\bibnamefont{Tabekoueng~Njitacke}},
  \bibinfo{author}{\bibfnamefont{I.}~\bibnamefont{Sami~Doubla}},
  \bibinfo{author}{\bibfnamefont{J.}~\bibnamefont{Kengne}}, \bibnamefont{and}
  \bibinfo{author}{\bibfnamefont{A.}~\bibnamefont{Cheukem}},
  \bibinfo{journal}{Chaos: An Interdisciplinary Journal of Nonlinear Science}
  \textbf{\bibinfo{volume}{30}}, \bibinfo{pages}{023101}
  (\bibinfo{year}{2020}{\natexlab{c}}).

\bibitem[{\citenamefont{Guo et~al.}(2020)\citenamefont{Guo, Zhu, Wang, and
  Ren}}]{ref34}
\bibinfo{author}{\bibfnamefont{Y.}~\bibnamefont{Guo}},
  \bibinfo{author}{\bibfnamefont{Z.}~\bibnamefont{Zhu}},
  \bibinfo{author}{\bibfnamefont{C.}~\bibnamefont{Wang}}, \bibnamefont{and}
  \bibinfo{author}{\bibfnamefont{G.}~\bibnamefont{Ren}},
  \bibinfo{journal}{Optik} \textbf{\bibinfo{volume}{218}},
  \bibinfo{pages}{164993} (\bibinfo{year}{2020}).

\bibitem[{\citenamefont{Joshi}(2021)}]{ref35}
\bibinfo{author}{\bibfnamefont{S.~K.} \bibnamefont{Joshi}},
  \bibinfo{journal}{IEEE Transactions on Circuits and Systems II: Express
  Briefs} \textbf{\bibinfo{volume}{69}}, \bibinfo{pages}{1737}
  (\bibinfo{year}{2021}).

\bibitem[{\citenamefont{Li et~al.}(2021{\natexlab{b}})\citenamefont{Li, Zhou,
  Wang, and Ma}}]{ref36}
\bibinfo{author}{\bibfnamefont{Z.}~\bibnamefont{Li}},
  \bibinfo{author}{\bibfnamefont{H.}~\bibnamefont{Zhou}},
  \bibinfo{author}{\bibfnamefont{M.}~\bibnamefont{Wang}}, \bibnamefont{and}
  \bibinfo{author}{\bibfnamefont{M.}~\bibnamefont{Ma}},
  \bibinfo{journal}{Nonlinear Dynamics} \textbf{\bibinfo{volume}{104}},
  \bibinfo{pages}{1455} (\bibinfo{year}{2021}{\natexlab{b}}).

\bibitem[{\citenamefont{Lin et~al.}(2020)\citenamefont{Lin, Wang, Sun, and
  Yao}}]{ref37}
\bibinfo{author}{\bibfnamefont{H.}~\bibnamefont{Lin}},
  \bibinfo{author}{\bibfnamefont{C.}~\bibnamefont{Wang}},
  \bibinfo{author}{\bibfnamefont{Y.}~\bibnamefont{Sun}}, \bibnamefont{and}
  \bibinfo{author}{\bibfnamefont{W.}~\bibnamefont{Yao}},
  \bibinfo{journal}{Nonlinear Dynamics} \textbf{\bibinfo{volume}{100}},
  \bibinfo{pages}{3667} (\bibinfo{year}{2020}).

\bibitem[{\citenamefont{Wu et~al.}(2020)\citenamefont{Wu, Ma, and
  Zhang}}]{ref38}
\bibinfo{author}{\bibfnamefont{F.}~\bibnamefont{Wu}},
  \bibinfo{author}{\bibfnamefont{J.}~\bibnamefont{Ma}}, \bibnamefont{and}
  \bibinfo{author}{\bibfnamefont{G.}~\bibnamefont{Zhang}},
  \bibinfo{journal}{Science China Technological Sciences}
  \textbf{\bibinfo{volume}{63}}, \bibinfo{pages}{625} (\bibinfo{year}{2020}).

\bibitem[{\citenamefont{Yao et~al.}(2021)\citenamefont{Yao, Zhou, Zhu, and
  Ma}}]{ref39}
\bibinfo{author}{\bibfnamefont{Z.}~\bibnamefont{Yao}},
  \bibinfo{author}{\bibfnamefont{P.}~\bibnamefont{Zhou}},
  \bibinfo{author}{\bibfnamefont{Z.}~\bibnamefont{Zhu}}, \bibnamefont{and}
  \bibinfo{author}{\bibfnamefont{J.}~\bibnamefont{Ma}},
  \bibinfo{journal}{Neurocomputing} \textbf{\bibinfo{volume}{423}},
  \bibinfo{pages}{518} (\bibinfo{year}{2021}).

\bibitem[{\citenamefont{Wouapi et~al.}(2020)\citenamefont{Wouapi, Fotsin,
  Louodop, Feudjio, Njitacke, and Djeudjo}}]{ref40}
\bibinfo{author}{\bibfnamefont{K.}~\bibnamefont{Wouapi}},
  \bibinfo{author}{\bibfnamefont{B.~H.} \bibnamefont{Fotsin}},
  \bibinfo{author}{\bibfnamefont{F.~P.} \bibnamefont{Louodop}},
  \bibinfo{author}{\bibfnamefont{K.~F.} \bibnamefont{Feudjio}},
  \bibinfo{author}{\bibfnamefont{Z.~T.} \bibnamefont{Njitacke}},
  \bibnamefont{and} \bibinfo{author}{\bibfnamefont{T.~H.}
  \bibnamefont{Djeudjo}}, \bibinfo{journal}{Cognitive Neurodynamics}
  \textbf{\bibinfo{volume}{14}}, \bibinfo{pages}{375} (\bibinfo{year}{2020}).

\bibitem[{\citenamefont{Wouapi et~al.}(2021)\citenamefont{Wouapi, Fotsin,
  Ngouonkadi, Kemwoue, and Njitacke}}]{ref41}
\bibinfo{author}{\bibfnamefont{M.~K.} \bibnamefont{Wouapi}},
  \bibinfo{author}{\bibfnamefont{B.~H.} \bibnamefont{Fotsin}},
  \bibinfo{author}{\bibfnamefont{E.~B.~M.} \bibnamefont{Ngouonkadi}},
  \bibinfo{author}{\bibfnamefont{F.~F.} \bibnamefont{Kemwoue}},
  \bibnamefont{and} \bibinfo{author}{\bibfnamefont{Z.~T.}
  \bibnamefont{Njitacke}}, \bibinfo{journal}{Cognitive neurodynamics}
  \textbf{\bibinfo{volume}{15}}, \bibinfo{pages}{315} (\bibinfo{year}{2021}).

\bibitem[{\citenamefont{Qin et~al.}(2021)\citenamefont{Qin, Menara, Bassett,
  and Pasqualetti}}]{qin2021phase}
\bibinfo{author}{\bibfnamefont{Y.}~\bibnamefont{Qin}},
  \bibinfo{author}{\bibfnamefont{T.}~\bibnamefont{Menara}},
  \bibinfo{author}{\bibfnamefont{D.~S.} \bibnamefont{Bassett}},
  \bibnamefont{and}
  \bibinfo{author}{\bibfnamefont{F.}~\bibnamefont{Pasqualetti}},
  \bibinfo{journal}{Physical Review Research} \textbf{\bibinfo{volume}{3}},
  \bibinfo{pages}{023218} (\bibinfo{year}{2021}).

\bibitem[{\citenamefont{Sysoeva et~al.}(2021)\citenamefont{Sysoeva, Sysoev,
  Prokhorov, Ponomarenko, and Bezruchko}}]{sysoeva2021reconstruction}
\bibinfo{author}{\bibfnamefont{M.~V.} \bibnamefont{Sysoeva}},
  \bibinfo{author}{\bibfnamefont{I.~V.} \bibnamefont{Sysoev}},
  \bibinfo{author}{\bibfnamefont{M.~D.} \bibnamefont{Prokhorov}},
  \bibinfo{author}{\bibfnamefont{V.~I.} \bibnamefont{Ponomarenko}},
  \bibnamefont{and} \bibinfo{author}{\bibfnamefont{B.~P.}
  \bibnamefont{Bezruchko}}, \bibinfo{journal}{Chaos, Solitons \& Fractals}
  \textbf{\bibinfo{volume}{142}}, \bibinfo{pages}{110513}
  (\bibinfo{year}{2021}).

\bibitem[{\citenamefont{Takembo et~al.}(2022)\citenamefont{Takembo, Fouda, and
  Kofane}}]{takembo2022modulational}
\bibinfo{author}{\bibfnamefont{C.~N.} \bibnamefont{Takembo}},
  \bibinfo{author}{\bibfnamefont{H.~P.~E.} \bibnamefont{Fouda}},
  \bibnamefont{and} \bibinfo{author}{\bibfnamefont{T.~C.}
  \bibnamefont{Kofane}}, \bibinfo{journal}{Indian Journal of Physics} pp.
  \bibinfo{pages}{1--9} (\bibinfo{year}{2022}).

\bibitem[{\citenamefont{Njitacke
  et~al.}(2022{\natexlab{b}})\citenamefont{Njitacke, Takembo, Awrejcewicz,
  Fouda, and Kengne}}]{njitacke2022hamilton}
\bibinfo{author}{\bibfnamefont{Z.~T.} \bibnamefont{Njitacke}},
  \bibinfo{author}{\bibfnamefont{C.~N.} \bibnamefont{Takembo}},
  \bibinfo{author}{\bibfnamefont{J.}~\bibnamefont{Awrejcewicz}},
  \bibinfo{author}{\bibfnamefont{H.~P.~E.} \bibnamefont{Fouda}},
  \bibnamefont{and} \bibinfo{author}{\bibfnamefont{J.}~\bibnamefont{Kengne}},
  \bibinfo{journal}{Chaos, Solitons \& Fractals}
  \textbf{\bibinfo{volume}{160}}, \bibinfo{pages}{112211}
  (\bibinfo{year}{2022}{\natexlab{b}}).

\bibitem[{\citenamefont{Willms et~al.}(2017)\citenamefont{Willms, Kitanov, and
  Langford}}]{Huy17}
\bibinfo{author}{\bibfnamefont{A.~R.} \bibnamefont{Willms}},
  \bibinfo{author}{\bibfnamefont{P.~M.} \bibnamefont{Kitanov}},
  \bibnamefont{and} \bibinfo{author}{\bibfnamefont{W.~F.}
  \bibnamefont{Langford}}, \bibinfo{journal}{R. Soc. Open Sci.}
  \textbf{\bibinfo{volume}{4}}, \bibinfo{pages}{170777} (\bibinfo{year}{2017}).

\bibitem[{\citenamefont{Kuramoto and
  Battogtokh}(2002)}]{kuramoto2002coexistence}
\bibinfo{author}{\bibfnamefont{Y.}~\bibnamefont{Kuramoto}} \bibnamefont{and}
  \bibinfo{author}{\bibfnamefont{D.}~\bibnamefont{Battogtokh}},
  \bibinfo{journal}{arXiv preprint cond-mat/0210694}  (\bibinfo{year}{2002}).

\bibitem[{\citenamefont{Sch{\"o}ll}(2016)}]{scholl2016synchronization}
\bibinfo{author}{\bibfnamefont{E.}~\bibnamefont{Sch{\"o}ll}},
  \bibinfo{journal}{Eur. Phys. J. Spec. Top.} \textbf{\bibinfo{volume}{225}},
  \bibinfo{pages}{891} (\bibinfo{year}{2016}).

\bibitem[{\citenamefont{Majhi et~al.}(2019)\citenamefont{Majhi, Bera, Ghosh,
  and Perc}}]{majhi2019chimera}
\bibinfo{author}{\bibfnamefont{S.}~\bibnamefont{Majhi}},
  \bibinfo{author}{\bibfnamefont{B.~K.} \bibnamefont{Bera}},
  \bibinfo{author}{\bibfnamefont{D.}~\bibnamefont{Ghosh}}, \bibnamefont{and}
  \bibinfo{author}{\bibfnamefont{M.}~\bibnamefont{Perc}},
  \bibinfo{journal}{Phys. Life Rev.} \textbf{\bibinfo{volume}{28}},
  \bibinfo{pages}{100} (\bibinfo{year}{2019}).

\bibitem[{\citenamefont{Omel’chenko}(2018)}]{omel2018mathematics}
\bibinfo{author}{\bibfnamefont{O.~E.} \bibnamefont{Omel’chenko}},
  \bibinfo{journal}{Nonlinearity} \textbf{\bibinfo{volume}{31}},
  \bibinfo{pages}{R121} (\bibinfo{year}{2018}).

\bibitem[{\citenamefont{Panaggio and Abrams}(2015)}]{panaggio2015chimera}
\bibinfo{author}{\bibfnamefont{M.~J.} \bibnamefont{Panaggio}} \bibnamefont{and}
  \bibinfo{author}{\bibfnamefont{D.~M.} \bibnamefont{Abrams}},
  \bibinfo{journal}{Nonlinearity} \textbf{\bibinfo{volume}{28}},
  \bibinfo{pages}{R67} (\bibinfo{year}{2015}).

\bibitem[{\citenamefont{Uhlhaas and Singer}(2006)}]{uhlhaas2006neural}
\bibinfo{author}{\bibfnamefont{P.~J.} \bibnamefont{Uhlhaas}} \bibnamefont{and}
  \bibinfo{author}{\bibfnamefont{W.}~\bibnamefont{Singer}},
  \bibinfo{journal}{Neuron} \textbf{\bibinfo{volume}{52}}, \bibinfo{pages}{155}
  (\bibinfo{year}{2006}).

\bibitem[{\citenamefont{Galinsky and Frank}(2021)}]{ref48}
\bibinfo{author}{\bibfnamefont{V.~L.} \bibnamefont{Galinsky}} \bibnamefont{and}
  \bibinfo{author}{\bibfnamefont{L.~R.} \bibnamefont{Frank}},
  \bibinfo{journal}{Physical review letters} \textbf{\bibinfo{volume}{126}},
  \bibinfo{pages}{158102} (\bibinfo{year}{2021}).

\bibitem[{\citenamefont{Njitacke
  et~al.}(2022{\natexlab{c}})\citenamefont{Njitacke, Fozin, Muni, Awrejcewicz,
  and Kengne}}]{njitacke2022energy}
\bibinfo{author}{\bibfnamefont{Z.~T.} \bibnamefont{Njitacke}},
  \bibinfo{author}{\bibfnamefont{T.~F.} \bibnamefont{Fozin}},
  \bibinfo{author}{\bibfnamefont{S.~S.} \bibnamefont{Muni}},
  \bibinfo{author}{\bibfnamefont{J.}~\bibnamefont{Awrejcewicz}},
  \bibnamefont{and} \bibinfo{author}{\bibfnamefont{J.}~\bibnamefont{Kengne}},
  \bibinfo{journal}{AEU-International Journal of Electronics and
  Communications} p. \bibinfo{pages}{154361}
  (\bibinfo{year}{2022}{\natexlab{c}}).

\bibitem[{\citenamefont{Roberts and Wessler}(1970)}]{ref49}
\bibinfo{author}{\bibfnamefont{L.~G.} \bibnamefont{Roberts}} \bibnamefont{and}
  \bibinfo{author}{\bibfnamefont{B.~D.} \bibnamefont{Wessler}}, in
  \emph{\bibinfo{booktitle}{Proceedings of the May 5-7, 1970, spring joint
  computer conference}} (\bibinfo{year}{1970}), pp. \bibinfo{pages}{543--549}.

\bibitem[{\citenamefont{Shu et~al.}(2021)\citenamefont{Shu, Zhou, Lian, Li,
  Zhao, Zeng, and Ma}}]{ref50}
\bibinfo{author}{\bibfnamefont{H.}~\bibnamefont{Shu}},
  \bibinfo{author}{\bibfnamefont{J.}~\bibnamefont{Zhou}},
  \bibinfo{author}{\bibfnamefont{Q.}~\bibnamefont{Lian}},
  \bibinfo{author}{\bibfnamefont{H.}~\bibnamefont{Li}},
  \bibinfo{author}{\bibfnamefont{D.}~\bibnamefont{Zhao}},
  \bibinfo{author}{\bibfnamefont{J.}~\bibnamefont{Zeng}}, \bibnamefont{and}
  \bibinfo{author}{\bibfnamefont{J.}~\bibnamefont{Ma}},
  \bibinfo{journal}{Nature Computational Science} \textbf{\bibinfo{volume}{1}},
  \bibinfo{pages}{491} (\bibinfo{year}{2021}).

\bibitem[{\citenamefont{Muni and Provata}(2020)}]{ref51}
\bibinfo{author}{\bibfnamefont{S.~S.} \bibnamefont{Muni}} \bibnamefont{and}
  \bibinfo{author}{\bibfnamefont{A.}~\bibnamefont{Provata}},
  \bibinfo{journal}{Nonlinear Dynamics} \textbf{\bibinfo{volume}{101}},
  \bibinfo{pages}{2509} (\bibinfo{year}{2020}).

\bibitem[{\citenamefont{Shepelev
  et~al.}(2020{\natexlab{a}})\citenamefont{Shepelev, Bukh, Muni, and
  Anishchenko}}]{shepelev2020role}
\bibinfo{author}{\bibfnamefont{I.}~\bibnamefont{Shepelev}},
  \bibinfo{author}{\bibfnamefont{A.}~\bibnamefont{Bukh}},
  \bibinfo{author}{\bibfnamefont{S.}~\bibnamefont{Muni}}, \bibnamefont{and}
  \bibinfo{author}{\bibfnamefont{V.}~\bibnamefont{Anishchenko}},
  \bibinfo{journal}{Chaos, Solitons \& Fractals}
  \textbf{\bibinfo{volume}{135}}, \bibinfo{pages}{109725}
  (\bibinfo{year}{2020}{\natexlab{a}}).

\bibitem[{\citenamefont{Shepelev
  et~al.}(2021{\natexlab{a}})\citenamefont{Shepelev, Muni, and
  Vadivasova}}]{shepelev2021spatiotemporal}
\bibinfo{author}{\bibfnamefont{I.}~\bibnamefont{Shepelev}},
  \bibinfo{author}{\bibfnamefont{S.}~\bibnamefont{Muni}}, \bibnamefont{and}
  \bibinfo{author}{\bibfnamefont{T.}~\bibnamefont{Vadivasova}},
  \bibinfo{journal}{Chaos: An Interdisciplinary Journal of Nonlinear Science}
  \textbf{\bibinfo{volume}{31}}, \bibinfo{pages}{043136}
  (\bibinfo{year}{2021}{\natexlab{a}}).

\bibitem[{\citenamefont{Shepelev
  et~al.}(2021{\natexlab{b}})\citenamefont{Shepelev, Muni, and
  Vadivasova}}]{shepelev2021synchronization}
\bibinfo{author}{\bibfnamefont{I.~A.} \bibnamefont{Shepelev}},
  \bibinfo{author}{\bibfnamefont{S.~S.} \bibnamefont{Muni}}, \bibnamefont{and}
  \bibinfo{author}{\bibfnamefont{T.~E.} \bibnamefont{Vadivasova}},
  \bibinfo{journal}{Chaos: An Interdisciplinary Journal of Nonlinear Science}
  \textbf{\bibinfo{volume}{31}}, \bibinfo{pages}{021104}
  (\bibinfo{year}{2021}{\natexlab{b}}).

\bibitem[{\citenamefont{Shepelev
  et~al.}(2020{\natexlab{b}})\citenamefont{Shepelev, Bukh, Muni, and
  Anishchenko}}]{shepelev2020quantifying}
\bibinfo{author}{\bibfnamefont{I.~A.} \bibnamefont{Shepelev}},
  \bibinfo{author}{\bibfnamefont{A.~V.} \bibnamefont{Bukh}},
  \bibinfo{author}{\bibfnamefont{S.~S.} \bibnamefont{Muni}}, \bibnamefont{and}
  \bibinfo{author}{\bibfnamefont{V.~S.} \bibnamefont{Anishchenko}},
  \bibinfo{journal}{Regular and Chaotic Dynamics}
  \textbf{\bibinfo{volume}{25}}, \bibinfo{pages}{597}
  (\bibinfo{year}{2020}{\natexlab{b}}).

\bibitem[{\citenamefont{Shepelev
  et~al.}(2021{\natexlab{c}})\citenamefont{Shepelev, Muni, Sch{\"o}ll, and
  Strelkova}}]{shepelev2021repulsive}
\bibinfo{author}{\bibfnamefont{I.~A.} \bibnamefont{Shepelev}},
  \bibinfo{author}{\bibfnamefont{S.~S.} \bibnamefont{Muni}},
  \bibinfo{author}{\bibfnamefont{E.}~\bibnamefont{Sch{\"o}ll}},
  \bibnamefont{and} \bibinfo{author}{\bibfnamefont{G.~I.}
  \bibnamefont{Strelkova}}, \bibinfo{journal}{Chaos: An Interdisciplinary
  Journal of Nonlinear Science} \textbf{\bibinfo{volume}{31}},
  \bibinfo{pages}{063116} (\bibinfo{year}{2021}{\natexlab{c}}).

\bibitem[{\citenamefont{Muni et~al.}(2022)\citenamefont{Muni, Fatoyinbo, and
  Ghosh}}]{muni2022dynamical}
\bibinfo{author}{\bibfnamefont{S.~S.} \bibnamefont{Muni}},
  \bibinfo{author}{\bibfnamefont{H.~O.} \bibnamefont{Fatoyinbo}},
  \bibnamefont{and} \bibinfo{author}{\bibfnamefont{I.}~\bibnamefont{Ghosh}},
  \bibinfo{journal}{arXiv preprint arXiv:2201.03219}  (\bibinfo{year}{2022}).

\bibitem[{\citenamefont{Njitacke~Tabekoueng
  et~al.}(2022)\citenamefont{Njitacke~Tabekoueng, Shankar~Muni, Fonzin~Fozin,
  Dolvis~Leutcho, and Awrejcewicz}}]{njitacke2022coexistence}
\bibinfo{author}{\bibfnamefont{Z.}~\bibnamefont{Njitacke~Tabekoueng}},
  \bibinfo{author}{\bibfnamefont{S.}~\bibnamefont{Shankar~Muni}},
  \bibinfo{author}{\bibfnamefont{T.}~\bibnamefont{Fonzin~Fozin}},
  \bibinfo{author}{\bibfnamefont{G.}~\bibnamefont{Dolvis~Leutcho}},
  \bibnamefont{and}
  \bibinfo{author}{\bibfnamefont{J.}~\bibnamefont{Awrejcewicz}},
  \bibinfo{journal}{Chaos: An Interdisciplinary Journal of Nonlinear Science}
  \textbf{\bibinfo{volume}{32}}, \bibinfo{pages}{053114}
  (\bibinfo{year}{2022}).

\end{thebibliography}
    
\end{document}